\documentclass[fleqn,usenatbib]{mnras}

\usepackage{newtxtext,newtxmath}

\usepackage[T1]{fontenc}

\DeclareRobustCommand{\VAN}[3]{#2}
\let\VANthebibliography\thebibliography
\def\thebibliography{\DeclareRobustCommand{\VAN}[3]{##3}\VANthebibliography}

\usepackage{graphicx}	
\usepackage{amsmath}	
\usepackage{orcidlink}

\DeclareRobustCommand{\ion}[2]{%
\relax\ifmmode
\ifx\testbx\f@series
{\mathbf{#1\,\mathsc{#2}}}\else
{\mathrm{#1\,\mathsc{#2}}}\fi
\else\textup{#1\,{\mdseries\textsc{#2}}}%
\fi}

\title[Anisotropic conduction on a moving mesh]{Anisotropic thermal conduction on a moving mesh for cosmological simulations}

\author[R. Y. Talbot et al.]{%
Rosie Y. Talbot$^{1}$\thanks{E-mail: rosie@mpa-garching.mpg.de (RYT)}\orcidlink{0000-0001-9393-7879}, 
Rüdiger Pakmor$^{1}$\orcidlink{0000-0003-3308-2420}, 
Christoph Pfrommer$^2$\orcidlink{0000-0002-7275-3998},  
Volker Springel$^1$\orcidlink{0000-0001-5976-4599}, 
Maria Werhahn$^1$\orcidlink{0000-0003-4984-4389}, 
\newauthor Rebekka Bieri$^3$\orcidlink{0000-0002-4554-4488}, 
Freeke van~de~Voort$^4$ \orcidlink{0000-0002-6301-638X}
\vspace*{1mm}\\%
$^{1}$Max-Planck-Institut f{\"u}r Astrophysik, Karl-Schwarzschild-Str. 1, D-85748 Garching, Germany\\%
$^{2}$Leibniz-Institute f\"{u}r Astrophysik Potsdam (AIP), An der Sternwarte 16, 14482 Potsdam, Germany\\%
$^{3}$Department of Astrophysics, University of Zurich, 8057 Zurich, Switzerland\\%
$^{4}$Cardiff Hub for Astrophysics Research and Technology, School of Physics and Astronomy, Cardiff University, Queen’s Buildings, Cardiff CF24 3AA, UK
}

\date{Accepted XXX. Received YYY; in original form ZZZ}

\pubyear{2024}

\begin{document}
\label{firstpage}
\pagerange{\pageref{firstpage}--\pageref{lastpage}}
\maketitle

\begin{abstract}
In weakly collisional, strongly magnetised plasmas such as the intracluster medium (ICM), hot accretion flows and the solar corona, the transport of heat and momentum occurs primarily along magnetic field lines. 
In this paper we present a new scheme for modelling anisotropic thermal conduction which we have implemented in the moving mesh code {\small AREPO}. Our implementation uses a semi-implicit time integration scheme which works accurately and efficiently with individual timestepping, making the scheme highly suitable for use in cosmological simulations. 
We apply the scheme to a number of test-problems including the diffusion of a hot patch of gas in a circular magnetic field, the progression of a point explosion in the presence of thermal conduction, and the evolution and saturation of buoyancy instabilities in anisotropically conducting plasmas. We use these idealised tests to demonstrate the accuracy and stability of the solver and highlight the ways in which anisotropic conduction can fundamentally change the behaviour of the system. Finally, we demonstrate the solver's capability when applied to highly non-linear problems with deep timestep hierarchies by performing high-resolution cosmological zoom-in simulations of a galaxy cluster with conduction. We show that anisotropic thermal conduction can have a significant impact on the temperature distribution of the ICM and that whistler suppression may be relevant on cluster scales.
The new scheme is, therefore, well suited for future work which will explore the role of anisotropic thermal conduction in a range of astrophysical contexts including the ICM of clusters and the circumgalactic medium of galaxies.
\end{abstract}




\section{Introduction}
Many astrophysical systems, such as hot accretion flows, the intracluster medium (ICM) of galaxy clusters, the solar wind and some phases of the interstellar medium (ISM), are weakly collisional and strongly magnetised. Such plasmas are characterised by a hierarchy of scales whereby the electron and ion gyroradii are much smaller than the electron mean free path of ion Coulomb collisions, which itself is smaller than characteristic lengthscale associated with the system.

As a result of this scale ordering, charged particles gyrate around magnetic field lines much faster than the rate at which they undergo Coulomb scattering. This effectively ties charged particles to field lines, meaning that the transport of heat and momentum occurs preferentially in the direction of the magnetic field and only gradients oriented along the magnetic field can be relaxed. Such systems are often well described by Braginskii magnetohydrodynamics \citep[MHD,][]{1965Braginskii} which differs from ideal MHD via the addition of anisotropic heat conduction and viscosity. 

Anisotropic heat conduction is thought to be important in a wide range of astrophysical processes in diverse contexts. For example: In determining the structure of the hot plasma in the solar corona \citep[see e.g.][]{1997YokoyamaShibata, 2011BingertPeter, 2013Bourdin+, 2020Ye+, 2022Navarro+} and in regulating energy transport in supernova remnants \citep[see e.g.][]{1975Chevalier, 2006Tilley+, 2008Balsara+, 2008Balsarab+}. Thermal conduction also affects the stability properties of the plasma in the formation and dynamics of multiphase structure in the ICM, the circumgalactic medium (CGM) and ISM \citep[see e.g.][]{2010Sharma+,2010Sharmab+, 2012ChoiStone, 2016BrueggenScannapieco, 2017Armillotta+, 2021JenningsLi, 2023BrueggenScannapieco}.

Thermal conduction is thought to play a significant role in shaping the properties and dynamics of the ICM. It may, for example, be important for distributing energy from active galactic nuclei (AGN) and offsetting radiative cooling in cool cores of galaxy clusters \citep{2003ZakamskaNarayan, 2011Voit,2015Voit+, 2016YangReynolds, 2017JacobPfrommerA, 2017JacobPfrommerB}. Anisotropic transport may also be relevant for explaining features such as cold fronts in the ICM \citep{2007MarkevitchVikhlinin, 2015ZuHone} which exhibit temperature changes on scales of order (or smaller than) the electron mean free path. Simulations find that magnetic fields tend to drape around cold fronts \citep{Lyutikov2006, Asai2007, 2008DursiPfrommer, 2010PfrommerDursi} which act to insulate them from the surroundings and potentially protect their coherence by suppressing mixing \citep{2003Fabian+, 2009DongStone, 2015ZuHone}.

Additionally, when heat transport occurs predominantly along the magnetic field, the stability properties of the plasma change markedly. The hydrodynamic stability to convection is no longer determined by the sign of the entropy gradient \citep{1958Schwarzschild} but rather by that of the temperature gradient, in combination with topology of the magnetic field. The resulting instabilities, driven by fast heat conduction, have the ability to affect the outer regions of galaxy clusters \citep[the magneto-thermal instability, MTI;][]{2000Balbus, 2001Balbus} and also the innermost regions of cool-core clusters \citep[the heat-flux-driven buoyancy instability, HBI;][]{2008Quataert}. The MTI generates a turbulent state driven by the background temperature gradient, consisting of density and velocity fluctuations across a broad range of scales. At saturation, the root mean square values of these fluctuations follow distinct power-law relationships with thermal diffusivity and the gradients of gas entropy and temperature \citep{Perrone2022a, Perrone2022b}.

Weakly collisional plasmas where the ratio of thermal-to-magnetic pressure is large ($\beta\sim 100$) are, however, susceptible to kinetic instabilities, driven by pressure anisotropies and heat fluxes \citep{2005Schekochihin+, 2014Kunz+, 2016Komarov+}. These microinstabilities act to enhance the scattering of charged particles, potentially leading to suppression of the conductivity below the collisional \citep{1962Spitzer} value \citep{2016Riquelme+,2021Berlok+}. Alternatively, conductivity can be suppressed below the classical Spitzer value in a weakly collisional plasma as a result of streaming electrons, which resonantly excite whistler waves. These waves frequently scatter electrons, causing a near-isotropisation of the electron distribution in the whistler wave frame. As a result, the mean electron transport speed along the magnetic field is reduced from its thermal value $\varv_\rmn{th}$ to the whistler phase speed $\sim \varv_\rmn{th}/\beta_\rmn{e}$, where $\beta_\rmn{e}$ is the electron plasma beta \citep{2016Roberg-Clark+, 2018Roberg-Clark+, 2018Komarov+, Drake2021}, thus, reducing the efficiency of heat conduction. While heat conduction is indeed suppressed for strong whistler suppression so that the MTI turbulence ceases to exist, externally-driven turbulence can revive the MTI turbulence and reestablish efficient heat conduction at a somewhat lower rate because of the intermittency of strong magnetic flux tubes along which most of the electron heat flux is transported \citep{Perrone2024b, Perrone2024a}.

Structure formation and evolution is a highly complex and non-linear problem with a vast range of relevant length- and time-scales that need to be considered. Numerical simulations therefore play a crucial role in exploring anisotropic thermal conduction in such contexts. Over the past two decades, a number of works have used simulations to investigate the effects of heat conduction in galaxy clusters \citep[see e.g.][]{2004Dolag+, Jubelgas2004, Parrish2010, 2010RuszkowskiOh, 2011RuszkowskiOh, 2016YangReynolds, 2017Kannan+, 2019Barnes+, 2019Su+, 2022Beckmann+, 2023Pellissier+}.

Some of these simulations were run with conduction solvers that employ explicit time-integration schemes which, for stability, require a timestep limit that scales with the square of the cell radius. Implicit or semi-implicit schemes are, therefore, preferable as they do not require this timestep criterion which can markedly reduce the attainable spatial resolution. 

When constructing such a scheme it is also important to ensure that energy is not allowed to flow from lower to higher temperatures, which can occur when not explicitly prevented. \citet{2007SharmaHammett} demonstrated, for a Cartesian mesh, how violating this entropy condition can be avoided with appropriate gradient limiters and \citet{2016Pakmor+} then generalised this procedure for an irregular mesh.

In this paper, we introduce a scheme for anisotropic thermal conduction which we have implemented into the moving-mesh code {\small AREPO} \citep{2010Springel, 2016PakmorNum, 2020Weinberger+}. Our scheme uses a semi-implicit time integration scheme, is compatible with individual timestepping and ensures the entropy condition is not violated. The underlying method is based on that of \citet{2016Pakmor+} which concerns itself with the problem of cosmic ray diffusion and is, itself, based on those of \citet{2007SharmaHammett} and \citet{2011SharmaHammett}. 

The anisotropic thermal conduction solver presented in this work is fundamentally different from the solver described in \citet{2016Kannan+}, which is also implemented in {\small AREPO}. Our solver represents an improvement over this earlier method, primarily due to the fact that it supports local timestepping; a feature which is crucial for efficiency and accuracy in cosmological simulations.

This paper is structured as follows. In Section~\ref{sec: equations}, we introduce the continuous form of the equations of Braginskii MHD and then describe our algorithm and its numerical implementation in Section~\ref{sec: implementation}. In Section~\ref{sec: Tests}, we assess the accuracy of our solver by performing several test problems including the diffusion of a hot patch of gas in a circular magnetic field, the progression of a point explosion in the presence of thermal conduction and the evolution and saturation of buoyancy instabilities in an anisotropically conducting plasma. In Section~\ref{sec: cluster}, we demonstrate the efficiency and stability of the solver when used in computationally demanding cosmological zoom simulations of a galaxy cluster and present some first results. Finally, in Section~\ref{sec: summary}, we summarise our results and provide a brief outlook.

\section{Basic equations}
\label{sec: equations}
The Braginskii MHD equations can be used to describe transport in a fully ionised, weakly collisional\footnote{By `weakly collisional' we are referring to systems where the gradient lengthscale is $10-10^3$ times larger than the Coulomb-collisional mean free path.} and strongly magnetised plasma. In such systems the electrons and ions are tied to magnetic field lines, resulting in the anisotropic transport of heat and momentum, governed by:
\begin{align}
    &\frac{\partial \rho}{\partial t} + \boldsymbol{\nabla}\boldsymbol{\cdot}(\rho \boldsymbol{\varv})=0\, ,\\
    &\frac{\partial \rho\boldsymbol{\varv}}{\partial t} + \boldsymbol{\nabla}\boldsymbol{\cdot}\bigg[\rho \boldsymbol{\varv}\boldsymbol{\varv}^{T} + P_{\rm tot}\mathbb{I} - \frac{\boldsymbol{B}\boldsymbol{B}^T}{4\pi}\bigg] = -\boldsymbol{\nabla}\boldsymbol{\cdot}\boldsymbol{\Pi}\, ,\\
    &\frac{\partial E}{\partial t} + \boldsymbol{\nabla}\boldsymbol{\cdot} \bigg[\big(E + P_{\rm tot}\big)\, \boldsymbol{\varv} - \frac{\boldsymbol{B}\big(\boldsymbol{\varv}\boldsymbol{\cdot}\boldsymbol{B}\big)}{4\pi}\bigg] = -\boldsymbol{\nabla}\boldsymbol{\cdot}(\boldsymbol{\Pi}\boldsymbol{\cdot}\boldsymbol{\varv}) -\boldsymbol{\nabla}\boldsymbol{\cdot} \boldsymbol{Q} \, , \label{eq: energy}\\
    &\frac{\partial \boldsymbol{B}}{\partial t} = \boldsymbol{\nabla} \times (\boldsymbol{\varv}\times\boldsymbol{B}) \, ,
\end{align}
where we have used Gaussian units. $\rho$, $\boldsymbol{\varv}$ and $\boldsymbol{B}$ are the local gas density, velocity, and magnetic field, respectively. $\mathbb{I}$ is the unit rank-two tensor and $P_{\rm tot}$ is the total pressure, accounting for thermal gas and magnetic fields 
\begin{equation}
    P_{\rm tot} = P + \frac{\boldsymbol{B}^2}{8\pi}\, .
\end{equation}
$E$ is the total energy density
\begin{equation}
    E = \rho u + \frac{1}{2} \rho \,\boldsymbol{\varv}^2 + \frac{\boldsymbol{B}^2}{8\pi} \, ,
\end{equation}
where $u$ is the specific internal energy. 
The anisotropic viscosity tensor $\boldsymbol{\Pi}$ is
\begin{equation}
    \boldsymbol{\Pi} = -\Delta P\,\bigg(\boldsymbol{b}\boldsymbol{b}^T - \frac{1}{3}\boldsymbol{\mathbb{I}}\bigg) \, ,
\end{equation}
where $\boldsymbol{b} = \boldsymbol{B}/|\boldsymbol{B}|$ and $\Delta P = P_{\perp} - P_{\parallel}$ is the pressure anisotropy, i.e.~the difference between the perpendicular and parallel pressures with respect to the magnetic field direction.  

The heat flow vector $\boldsymbol{Q}$ is given by
\begin{equation}
\label{eq: aniso j}
	\boldsymbol{Q} = -\chi\big[\boldsymbol{b}(\boldsymbol{b}\boldsymbol{\cdot}\boldsymbol{\nabla} T)\big] \, ,
\end{equation}
where $\chi$ is the conductivity along the magnetic field and $T$ is the temperature, which is related to the specific internal energy via $u = c_{\rm v} \,T$. The specific heat capacity at constant volume is given by $c_{\rm v} = k_\rmn{B} / [(\gamma-1)\mu\, m_\mathrm{p}]$, where $k_\mathrm{B}$ is the Boltzmann constant, $\mu$ is the mean molecular weight, $m_\mathrm{p}$ is the proton mass and $\gamma$ is the adiabatic index.

In a collisional theory of transport processes, the diffusive transfer of heat is dominated by electrons and mediated by particle-particle Coulomb collisions. The resulting conductivity is often referred to as the `Spitzer conductivity' \citep{1962Spitzer} and is given by
\begin{equation}
\label{eq: spitzer}
	\chi_{\rm sp} =  1.84\times 10^{-5} \, \frac{T^{5/2}}{\ln C} \; {\rm erg \, s^{-1} \, K^{-1} \, cm^{-1}} \, ,
\end{equation}
where the temperature, $T$, is measured in Kelvin and $\ln C\approx37$ is the Coulomb logarithm. 

High $\beta$, weakly collisional plasmas, however, are susceptible to a variety of kinetic instabilities which act to alter the mean electron transport velocity which may, ultimately, lead to a suppression of electron transport. 

At saturation, the whistler instability has been found to establish a marginal heat flux which is suppressed by a factor of $1/\beta_\rmn{e}$, the inverse of the electron plasma beta \citep{2018Roberg-Clark+, 2018Komarov+}. To account for this, one can assume a functional form for the conductivity \citep{2018Komarov+} that smoothly interpolates between the two regimes
\begin{equation}
\label{eq: kappa interp}
	\chi_{\rm sat, whist} = \frac{\chi_{\rm sp}}{1 + (1/3) \,\beta_\rmn{e} \, \lambda_{\rm mfp, e} \, / \, l_{\rm T, \parallel}} \, ,
\end{equation}
where $l_{\rm T, \parallel} = |\boldsymbol{b} \boldsymbol{\cdot} \boldsymbol{\nabla} \ln T|^{-1}$ is the temperature gradient scale parallel to the magnetic field and $\lambda_{\rm mfp, e}$ is the electron mean free path, given by
\begin{equation}
	\lambda_{\rm mfp, e} = \frac{3^{3/2} k_\rmn{B}^2 T^2}{4 \pi^{1/2}n_{\rm e}e^4 \ln C} \, ,
\end{equation}
where $n_{\rm e}$ is the electron number density and $e$ is the electron charge.

One can also construct an analogous form for the conductivity that incorporates the saturation of the heat flux due to the free-streaming of electrons \citep{1977CowieMcKee}
\begin{equation}
\label{eq: kappa interp free}
	\chi_{\rm sat, free} = \frac{\chi_{\rm sp}}{1 + 4.2 \, \lambda_{\rm mfp, e} \, / \, l_{\rm T, \parallel}} \, .
\end{equation} 
In the outskirts of galaxy clusters, however, where the plasma beta is expected to be high ($\beta\gtrsim 100$), whistler suppression is likely to be more relevant. 


\section{Implementation} 
\label{sec: implementation}

We now describe how we implemented the anisotropic thermal conduction solver into the {\small AREPO} code \citep{2010Springel, 2016PakmorNum, 2020Weinberger+}. Note that there is already a first version of an anisotropic conduction solver implemented in {\small AREPO} \citep{2016Kannan+}. The solver presented in this work, however, is fundamentally different and is compatible with local timestepping. We discuss these differences in more detail in Section~\ref{subsec: local timestepping}. Note also that a solver for Braginskii viscosity has already been implemented in {\small AREPO} by \citet{2020Berlok+}. 

{\small AREPO} solves the equations of ideal MHD on an unstructured Voronoi mesh using a second order finite volume scheme \citep{2011Pakmor+, 2013PakmorSpringel}. The Voronoi mesh is constructed from a set of mesh-generating points that can move with arbitrary velocities, but which are typically set to the local fluid velocity, resulting in quasi-Lagrangian behaviour. {\small AREPO} computes self-gravity using a tree-PM method and couples it to MHD with a second-order Leapfrog scheme \citep{2010Springel, 2021Springel+}.

The anisotropic thermal conduction scheme, described in this work, is based on that of the cosmic ray diffusion solver presented in \citet{2016Pakmor+}. This approach, itself, generalises and extends the flux limiting scheme of \citet{2007SharmaHammett} and the semi-implicit time-integration scheme of \citet{2011SharmaHammett} for use with unstructured meshes and local timestepping. 

We have implemented this solver in such a way that it can be used for both cosmological and non-cosmological simulations. For simplicity, however, we will continue below to use physical coordinates to describe the scheme and explain how to implement the comoving form of the equations in Section~\ref{subsec: comoving}.

We now focus on the treatment of just the conduction term in equation~(\ref{eq: energy}),
\begin{equation}
\label{eq: atc}
	\frac{\partial u}{\partial t} = \frac{1}{\rho c_{\rm v}}\, \boldsymbol{\nabla}\boldsymbol{\cdot} \bigg[ \chi \, \boldsymbol{b}(\boldsymbol{b}\boldsymbol{\cdot}\nabla u)\bigg] \, ,
\end{equation}
where we have re-cast the equation in terms of the specific internal energy, assuming $c_{\rm v}$ is spatially constant. 

Our numerical treatment of thermal conduction necessarily differs from that of cosmic ray diffusion, described in \citet{2016Pakmor+}, in a number of ways. Firstly, the treatment of cosmic ray diffusion assumes that the diffusivity is spatially and temporally constant and that $\chi$, therefore, commutes with the gradient operator. This is, however, not a good assumption for thermal conduction and, therefore, requires an additional treatment. Note, however, that this also may not be a good assumption for cosmic ray diffusion \citep[see e.g.][]{2023Thomas+}. Some of the methods described in this paper may, therefore, be relevant for future one-moment treatments of cosmic ray diffusion. In addition to this, there is an extra factor of $1/\rho$ before the flux term in the thermal conduction equation (equation~\ref{eq: atc}), meaning that extra care has to be taken in the integration to ensure the scheme is conservative.

\subsection{Spatial discretisation}
We now spatially discretise equation~(\ref{eq: atc}) and begin by integrating over volume
\begin{equation}
	V\frac{\partial u}{\partial t} = \frac{1}{\rho c_{\rm v}}\int_{V} \boldsymbol{\nabla}\boldsymbol{\cdot} \bigg[ \chi \, \boldsymbol{b}(\boldsymbol{b}\boldsymbol{\cdot}\boldsymbol{\nabla} u)\bigg]{\rm d}V \, ,
\end{equation}
where we have assumed that $\rho$ and $u$ are constant throughout the volume and that $\int{\rm d}V$ commutes with the time derivative. 
Using Gauss' theorem then gives
\begin{equation}
	\frac{\partial u}{\partial t} = \frac{1}{V\rho c_{\rm v}}\int_{\partial V} \bigg[ \chi \, \boldsymbol{b}(\boldsymbol{b}\boldsymbol{\cdot}\boldsymbol{\nabla} u)\bigg]\boldsymbol{\cdot} {\rm d}\boldsymbol{A} \, ,
\end{equation}
where ${\rm d}\boldsymbol{A}$ is the area element on the surface $\partial V$, directed along the outward normal.

We now take the integral to be over one cell in the simulation, indexed by $i$,
\begin{equation}
\label{eq: spatial disc}
	\frac{\partial u_i}{\partial t} = \frac{1}{m_i c_{\rm v}}\sum_{j} \bigg[ \chi_{ij} \, (\boldsymbol{b}_{ij}\boldsymbol{\cdot}\boldsymbol{\nabla} u_{ij}) \, (\boldsymbol{b}_{ij}\boldsymbol{\cdot}\boldsymbol{A}_{ij})\bigg] \, ,
\end{equation}
where the sum is taken over all faces of the cell, and quantities indexed by $ij$ are calculated in the interface between cells $i$ and $j$ which has area $\boldsymbol{A}_{ij}$. Note that we have also used the relation $\rho_i = m_i / V_i$, where $m_i$ and $V_i$ are the mass and volume of the cell, respectively. 

We have also implemented a solver for \textit{isotropic} conduction. In this case, the discretised equation that determines the evolution of the specific internal energy is
\begin{equation}
\label{eq: spatial disc iso}
	\frac{\partial u_i}{\partial t} = \frac{1}{m_i c_{\rm v}}\sum_{j} \bigg[ \chi_{ij}\,  (\boldsymbol{\nabla} u_{ij}\boldsymbol{\cdot}\boldsymbol{A}_{ij})\bigg] \, .
\end{equation}

\subsubsection{Estimating quantities in the interface}
\label{subsub: interface}
To evaluate the flux on the right-hand side of equation~(\ref{eq: spatial disc}), the gradient of the specific thermal energy, $\boldsymbol{\nabla} u_{ij}$, and the magnetic field direction, $\boldsymbol{b}_{ij}$, in the interfaces of all cells need to be determined. 

When calculating the gradient of the thermal energy in the interface, care has to be taken to ensure the resulting solution is physical. When not explicitly prevented in the gradient estimates, the solution may lead to heat flowing from a cold cell to a hot cell. This can, however, be avoided by using gradient limiters, as described in \citet{2007SharmaHammett}. To calculate the gradients we largely follow the procedure described in \citet{2016Pakmor+} which generalises the procedure of \citet{2007SharmaHammett} for a moving mesh. We now briefly summarise the relevant details here. 

To estimate the specific thermal energy gradient in the interface, we first determine the gradients at the corners of the Voronoi face. The interface estimate is then built from these corner estimates. In (2D) 3D, every corner of a Voronoi cell corresponds to the centre of the (circumcircle) circumsphere of a (triangle) tetrahedron in the dual Delaunay tessellation. The corners of a Delaunay (triangle) tetrahedron correspond to mesh-generating points in the Voronoi mesh, so every corner has (three) four adjacent cells. 

To estimate the specific thermal energy gradient at the corner, we perform a least-squares fit, using the values of the specific thermal energy at the centres of mass\footnote{Note that the centre of mass of a Voronoi cell is not necessarily spatially coincident with the associated mesh-generating point.} of these adjacent cells \citep[see section $2.1$ of][for an in-depth explanation]{2016Pakmor+}. 

\begin{figure}
    \centering
    \includegraphics[width=0.49\textwidth]{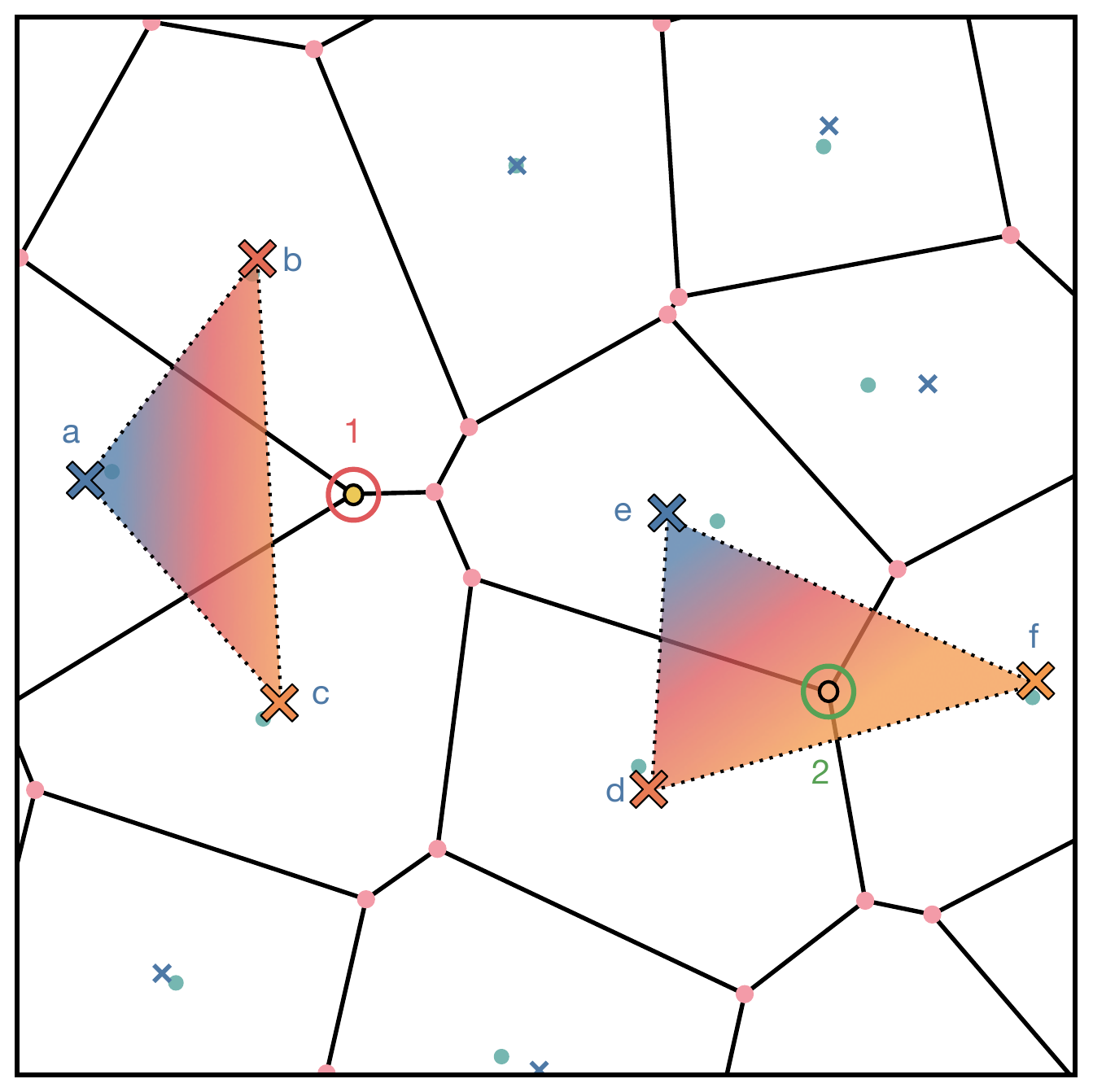}
    \caption[Estimating quantities at corners]
    {A sketch illustrating the gradient estimate at the corners of the Voronoi mesh in 2D. Mesh generating points are indicated by turquoise circles and the centres of mass of cells by blue crosses. The corners of the mesh are indicated by pink circles and are connected by the faces that form the Voronoi mesh, shown in black. A quantity at corner $1$ (highlighted by the red circle) is estimated from the values at the centres of mass $a$, $b$ and $c$ of the three adjacent cells. The values at these three points are indicated by the colouring of each cross and the shading within the triangle corresponds to the least-squares estimate of the gradient at the corner $1$. Since the corner, $1$, lies outside of the triangle, the value predicted at this point (illustrated by its colouring) is larger than those at $a$, $b$ and $c$ and has been extrapolated, producing a new extremum. The gradient at corner $2$ (highlighted by the green circle), however, is calculated from the values at the centres of mass $d$, $e$ and $f$. This corner lies within the triangle $def$ and the value predicted by the least-squares fit lies within the range of values at $d$, $e$ and $f$, corresponding to an interpolation.}
    \label{fig: corner estimate}
\end{figure}

If a corner lies outside of the (triangle) tetrahedron formed by the centres of mass of the (three) four neighbouring cells the gradient estimate will be an extrapolation rather than an interpolation. This is illustrated in Fig.~\ref{fig: corner estimate} where, for simplicity, we consider the 2D Voronoi mesh. The value of a quantity and its gradient are estimated at corner $1$ from the values at the centres of mass, $a$, $b$ and $c$, of the three adjacent cells. In the figure, the values at these three points are indicated by the colouring of the crosses and the shading within the triangle corresponds to the fit to the gradient at corner $1$. As corner $1$ lies outside of the triangle $abc$, the predicted value (indicated by its colouring) is larger than those at points $a$, $b$ and $c$, and has been extrapolated. The gradient at corner $2$, on the other hand, is estimated from the values at the centres of mass $d$, $e$ and $f$ of its three adjacent cells. This corner lies within the triangle $def$ and the value predicted by the least-squares fit is within the range of those at $d$, $e$ and $f$, corresponding to an interpolation.

We mark corners where extrapolation has occurred as `problematic' and use a different treatment for the contribution of this corner to gradient estimates in the interface; we will describe this alternative treatment shortly.

\begin{figure*}
    \centering
    \includegraphics[width=\textwidth]{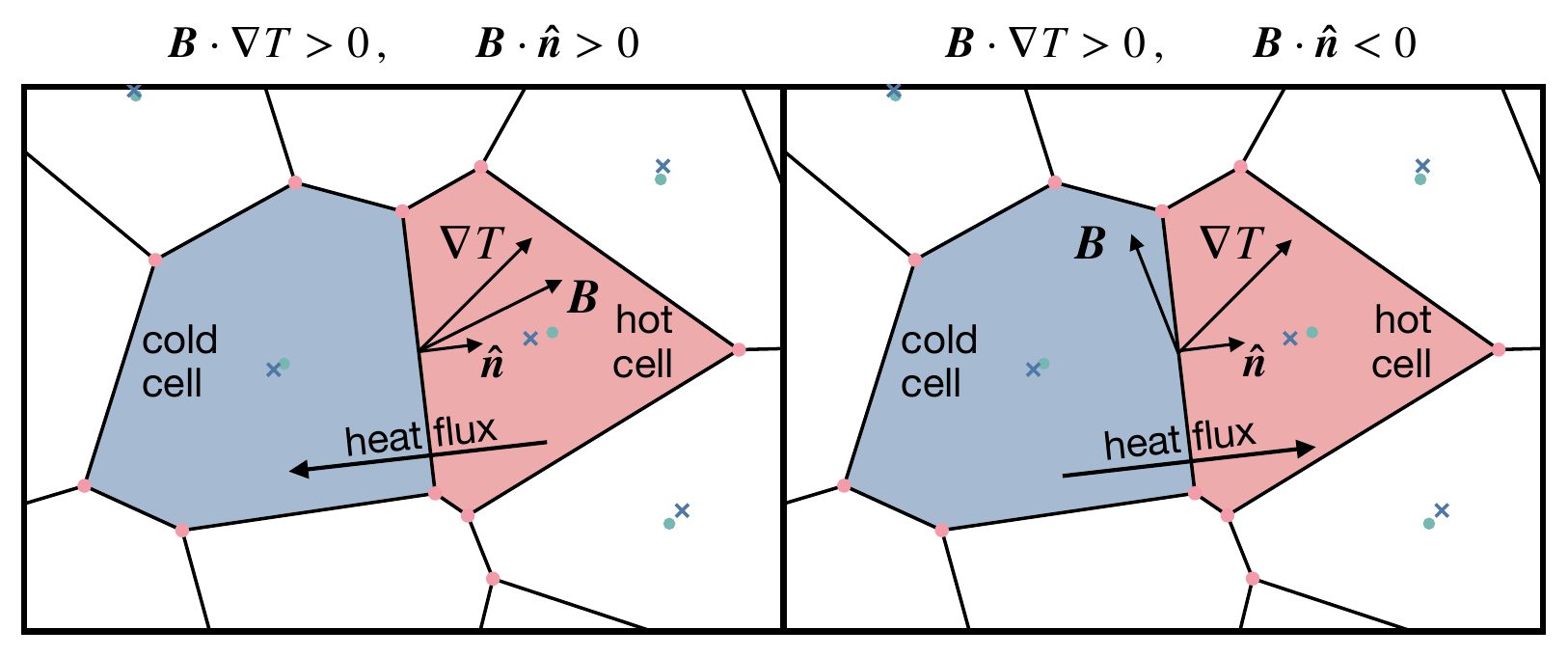}
    \caption[Flux direction]
    {A sketch illustrating how anisotropic diffusion schemes are prone to violations of the entropy criterion. Mesh generating points are indicated by turquoise circles and the centres of mass of cells by blue crosses. The corners of the mesh are indicated by pink circles and are connected together by the faces that form the Voronoi mesh, shown in black. Both panels show the same configuration of the Voronoi mesh, with the only difference being the direction of the magnetic field estimate in the interface (as indicated by the arrow labelled $\boldsymbol{B}$). On the left-hand side, the configuration is such that the projection of the temperature gradient estimate onto the magnetic field direction and the projection of the magnetic field onto the normal to the face are positive. In this case, heat will flow across this face from the hot cell to the cold (see equation~\ref{eq: spatial disc}). In the right hand panel, however, the projection of the magnetic field onto the face normal is now negative, meaning that heat will flow across this face from the cold cell to the hot. This is a result of the numerical discretisation of the diffusion problem and can lead to extrema being accentuated.}
    \label{fig: entropy}
\end{figure*}

In the calculation of the fluxes, we split the gradient of the thermal energy into normal $\boldsymbol{\nabla} u_{ij, {\rm N}}$ and tangential $\boldsymbol{\nabla} u_{ij, {\rm T}}$ components relative to the normal to the interface. 
\begin{equation}
	\boldsymbol{\nabla} u_{ij} = \boldsymbol{\nabla} u_{ij, {\rm N}} + \boldsymbol{\nabla} u_{ij, {\rm T}} \, .
\end{equation}
The contributions to the flux from these two components are then calculated separately and added together to get the total flux across the interface. 

In Fig.~\ref{fig: entropy} we show a sketch illustrating why anisotropic diffusion schemes are prone to violations of the entropy criterion. Both panels show exactly the same configuration of the Voronoi mesh, with the only difference being the direction of the magnetic field estimate in the interface. In the panel on the left, the quantity $(\boldsymbol{B\cdot\nabla T})(\boldsymbol{B\cdot\hat{n}})$ is positive (where $\boldsymbol{\hat{n}}$ is the normal to the face). Inspecting equation~(\ref{eq: spatial disc}) shows that, in this configuration, heat will flow across the face, from the hot cell to the cold. In the right hand panel, however, the projection of the magnetic field onto the face normal is now negative, meaning that heat will flow across this face in the opposite direction. 

To ensure that temperature extrema are not accentuated we use a procedure very similar to that described in \citet{2007SharmaHammett} which requires two steps. First, when estimating the normal gradient of the thermal energy in the interface (described below), we ensure that energy associated with the normal flux flows across a face from the hotter cell to the colder. Second, we slope-limit the tangential gradient, $\boldsymbol{\nabla} u_{ij, {\rm T}}$, using a generalised version of the van Leer limiter \citep{1984vanLeer}
\begin{equation}
\label{eq: VLL}
     \boldsymbol{\nabla} u_{ij, {\rm T}} = \frac{N}{\sum_{k}(\boldsymbol{\nabla} u_{k, {\rm T}})^{-1}} \, ,
\end{equation}
where the sum runs over all $N$ corners (indexed by $k$) of the face $ij$. If the tangential gradients at the corners $\boldsymbol{\nabla} u_{k, {\rm T}}$ do not all have the same sign, the tangential gradient in the interface is set to zero.

We calculate the normal component of the gradient in the interface by taking an average of the values at the corner, weighted by the fractional area of the Voronoi face that is closest to this corner. This alone does not, however, guarantee that the result satisfies the entropy condition. We, therefore, additionally calculate a `simple' finite difference estimate using the values of the specific thermal energy at the centres of mass of the two cells on either side of the interface. While the corner averaged normal gradient estimate has a larger stencil, this `simple' estimate is more robust. We then compare the signs of these two gradient estimates and, if they disagree, we use the `simple' finite difference estimate which satisfies the entropy condition by construction. We, additionally, replace the contribution of corners that were previously flagged as `problematic' (due to extrapolation in the gradient estimate at the corner) with this `simple' estimate.

The magnetic field direction in the interface, $\boldsymbol{b}_{ij}$, is calculated in a similar way. We use a least-squares fit to get a first estimate at the corners, and then calculate the interface value by taking a weighted average of these corner estimates\footnote{The weights are the same as those described above for the thermal energy gradient in the interface, i.e. the fractional area of the Voronoi face that is closest to the relevant corner.}. We then calculate $c_{\rm v}$ assuming the gas is fully ionised with primordial composition. Next, we calculate the conductivity in the interface, $\chi_{ij}$, by taking the arithmetic average of the conductivities of the two cells on either side of the interface
\begin{equation}
\label{eq: arithmetic}
    \chi_{ij} = \frac{1}{2}(\chi_i + \chi_j) \, .
\end{equation}

For the scheme outlined in \citet{2007SharmaHammett} they suggest instead using the harmonic average for numerical stability. In our scheme, however, we do not observe any stability problems when using the arithmetic average. Additionally, in Appendix~\ref{app: avg}, we use point explosion simulations (analogous to those presented in Section~\ref{subsub: sedov iso}) to explore the effects of these two averaging procedures and show that, for our scheme, the arithmetic mean best reproduces the analytic predictions of the speed of advance of the conduction fronts.

Note that, when using the functional form of the conductivity that interpolates between the whistler suppressed and collisional regimes (see equation~\ref{eq: kappa interp}), we use the unlimited temperature gradient to calculate the temperature gradient lengthscale. Typically, {\small AREPO} stores the limited gradients that are used in the flux calculations, which can be significantly lower than the true gradients in regions where the gradients are not well resolved. Using these limited gradients in the calculation of $l_{\rm T,\parallel}$ could lead to underestimates of the level of suppression. Additionally, since {\small AREPO} uses the single-fluid MHD approximation we do not have access to the electron temperature. When calculating the whistler suppression factor we, therefore, use the bulk plasma temperature to calculate the plasma beta.

For the case of isotropic conduction, only the gradient of the specific thermal energy normal to the surface contributes to the flux, which we estimate in the same way as for $\boldsymbol{\nabla} u_{ij, {\rm N}}$.

\subsection{Time integration}

We implement three different time integration schemes for the anisotropic thermal conduction solver: an explicit solver, a semi-implicit, linear solver and a semi-implicit, non-linear solver. Note that in the descriptions of these methods in the next three sub-sections, the equations are strictly only valid when using global timesteps. In Section~\ref{subsec: local timestepping} we discuss the adaptations that need to be made when using local timestepping.

\subsubsection{Explicit time integration}
\label{subsubsec: explicit time integration}
For the explicit integration scheme, we evolve the specific thermal energy using two half timesteps updates which are carried out immediately before and after the first and second gravity kicks, respectively:
\begin{align}
\label{eq: Explicit}
	u_i^{n + 1/2} &= u_i^{n} +\frac{\Delta t}{2 m_i^n c_{\rm v}}\sum_{j} \bigg[ \chi_{ij}^n \, \big(\boldsymbol{b}_{ij}\boldsymbol{\cdot}\boldsymbol{\nabla} u_{ij,{\rm T}}^n\big)\,(\boldsymbol{b}_{ij}\boldsymbol{\cdot}\boldsymbol{A}_{ij})\bigg] \nonumber \\ 
    & + \frac{\Delta t}{2 m_i^n c_{\rm v}}\sum_{j} \bigg[ \chi_{ij}^n \, \big(\boldsymbol{b}_{ij}\boldsymbol{\cdot}\boldsymbol{\nabla} u_{ij,{\rm N}}^n\big) \,(\boldsymbol{b}_{ij}\boldsymbol{\cdot}\boldsymbol{A}_{ij})\bigg]\nonumber ,\\
    u_i^{n+1}\hspace{3pt} &= u_i^{n+1/2} + \frac{\Delta t}{2 m_i^{n+1} c_{\rm v}}\sum_{j} \bigg[ \chi_{ij}^{n+1/2} \, \big(\boldsymbol{b}_{ij}\boldsymbol{\cdot}\boldsymbol{\nabla} u_{ij,{\rm T}}^{n+1/2}\big)\,(\boldsymbol{b}_{ij}\boldsymbol{\cdot}\boldsymbol{A}_{ij})\bigg] \nonumber \\ 
    & + \frac{\Delta t}{2 m_i^{n+1} c_{\rm v}}\sum_{j} \bigg[\chi_{ij}^{n+1/2} \, \big(\boldsymbol{b}_{ij}\boldsymbol{\cdot}\boldsymbol{\nabla} u_{ij,{\rm N}}^{n+1/2}\big)\,(\boldsymbol{b}_{ij}\boldsymbol{\cdot}\boldsymbol{A}_{ij})\bigg]\, ,
\end{align}
where the superscripts $n$, $n+1/2$, and $n+1$ correspond to the beginning, mid-point and end of a timestep of length $\Delta t$, respectively. We use the same procedure for the explicit treatment of isotropic conduction, with the appropriate flux term (equation~\ref{eq: spatial disc iso}).

For stability, this explicit integration scheme requires a timestep constraint of the form
\begin{equation}
\label{eq: tstp}
    \Delta t < \eta \frac{(\Delta x)^2}{\kappa} \, ,
\end{equation}
where $\Delta x$ is the cell diameter, $\eta < 1$ is a constant, and 
\begin{equation}
    \kappa = \frac{\chi}{c_{\rm v} \rho} \, ,
\end{equation}
is the thermal diffusivity. When determining this timestep for each cell, we use the maximum value of $\chi$ in any of its interfaces to calculate $\kappa$. 

Note that this timestep criterion can become prohibitively restrictive due to its quadratic dependence on the cell size, severely limiting the achievable spatial resolution. It is therefore preferable to use semi-implicit or implicit time-integration schemes, which do not require this timestep criterion for stability. 

\subsubsection{Semi-implicit, linear solver}
\label{subsubsec: linear solver}
The gradient limiter that ensures the entropy condition is not violated (equation~\ref{eq: VLL}) introduces non-linearities into the tangential flux estimate, which significantly increase the numerical complexity when using an implicit integrator.
The normal gradient estimate on the other hand has an explicit, linear dependence on internal energy. But it has an additional, implicit dependence on temperature via the conductivity (which is often modelled as being temperature dependent). In this `semi-implicit, linear solver' we assume $\chi_{ij}$ for each interface is constant during the determination of the normal flux. In the following section we discuss an integration scheme that relaxes this assumption. 

When changes to $\chi_{ij}$ across a timestep are ignored, a semi-implicit scheme can be formulated which is almost as stable as a fully implicit scheme and in which only one linear implicit problem is solved per timestep \citep[see][]{2011SharmaHammett, 2016Pakmor+}. 
To this end we split the calculation into two parts. First, the flux associated with the tangential component of the specific thermal energy gradient is evolved using a forward-Euler method,
\begin{equation}
	u_i^{\tilde{n}} = u_i^{n} +\frac{\Delta t}{m_i^{n+1} c_{\rm v}}\sum_{j} \bigg[\chi_{ij}^n \, \big(\boldsymbol{b}_{ij}\boldsymbol{\cdot}\boldsymbol{\nabla} u_{ij,{\rm T}}^n\big)\,(\boldsymbol{b}_{ij}\boldsymbol{\cdot}\boldsymbol{A}_{ij})\bigg] \, ,
\end{equation}
where the superscript $\tilde{n}$ indicates intermediate values. In the second step, the thermal energies are advanced due to the flux associated with the normal component of the specific thermal energy gradient using an implicit backward-Euler scheme,
\begin{equation}
	u_i^{n+1} = u_i^{\tilde{n}} + \frac{\Delta t}{m_i^{n+1} c_{\rm v}}\sum_{j} \bigg[\chi_{ij}^n \, \big(\boldsymbol{b}_{ij}\boldsymbol{\cdot}\boldsymbol{\nabla} u_{ij,{\rm N}}^{n+1}\big)\,(\boldsymbol{b}_{ij}\boldsymbol{\cdot}\boldsymbol{A}_{ij}) \bigg]\, .
\end{equation}

To solve this linear system of equations, we use the {\small HYPRE} library \citep{2002FalgoutYang} and carry out a two-step procedure, first using a generalised minimal residual (GMRES) solver \citep{1986SaadSchultz} iteratively until the residual drops below $10^{-8}$. If this condition is not met within $200$ iterations, we add an algebraic multigrid preconditioner \citep{2002HensonYang} to GMRES which is then iterated until the residual drops below $10^{-8}$.

We apply a flux limiter to the first (explicit) part of this calculation to ensure that cells are not completely drained of thermal energy during the exchange. Such a limiter is, however, not required in the second (implicit) part of the calculation. 

We find that this scheme does not require a timestep criterion similar to that required by explicit schemes (see equation~\ref{eq: tstp}) and is stable on much longer timesteps, similar to what was observed in \citet{2011SharmaHammett} and \citet{2016Pakmor+}. The implicit part of the flux calculation is unconditionally stable. As the timesteps become very long, however, the explicit fluxes may become so large that the flux limiter activates and the solution becomes more isotropic. We, however, do not observe such behaviour when cells are integrated on their MHD timestep.

We also treat the case of isotropic conduction similarly and split the flux into two steps of size $\Delta t/2$. We integrate the first step using the explicit scheme and the second using the implicit scheme. This semi-implicit scheme is fully second-order convergent in the case of isotropic conduction \citep{2016Pakmor+}. We could in principle treat isotropic conduction purely implicitly (as only normal gradient estimates are required which are linear in the specific internal energy of the cells), though this scheme would only be first-order accurate in time.

\subsubsection{Semi-implicit, non-linear solver}
\label{subsub: non linear solver}
When the conductivity has a temperature dependence (e.g. Spitzer), the fluxes of thermal energy associated with conduction will necessarily lead to changes in the conductivity. To test whether taking into account changes in $\chi_{ij}$ across each timestep affects our results, we also implemented a `non-linear solver'. 

Put simply, this `semi-implicit, non-linear solver' iteratively applies the procedure described in the previous section for the `semi-implicit, linear solver' and updates the conductivity and gradients at each iteration. Specifically, we iteratively calculate
\begin{align}
	u_i^{\tilde{n},m} &= u_i^{n,0} + \frac{\Delta t}{m_i^{n+1} c_{\rm v}}\sum_{j} \bigg[ \chi_{ij}^{n,m-1} \, \big(\boldsymbol{b}_{ij}\boldsymbol{\cdot}\boldsymbol{\nabla} u_{ij,{\rm T}}^{n,m-1}\big)\,(\boldsymbol{b}_{ij}\boldsymbol{\cdot}\boldsymbol{A}_{ij})\bigg] \label{eq nonlin t}\, ,\\
	u_{i}^{n,m*} &= u_i^{\tilde{n},m} + \frac{\Delta t}{m_i^{n+1} c_{\rm v}}\sum_{j} \bigg[ \chi_{ij}^{\tilde{n},m} \, \big(\boldsymbol{b}_{ij}\boldsymbol{\cdot}\boldsymbol{\nabla} u_{ij,{\rm N}}^{n,m*}\big)\,(\boldsymbol{b}_{ij}\boldsymbol{\cdot}\boldsymbol{A}_{ij})\bigg] \label{eq nonlin n}\, ,
\end{align}
where $n$ indexes the timestep and $m$ indexes the iteration. 
$\tilde{n}$ denotes an intermediate quantity, after the tangential flux has been evolved but before that of the normal flux. We use the intermediate thermal energies to calculate the conductivity, $\chi_{ij}^{\tilde{n},m}$, that is used in the normal flux (equation~\ref{eq nonlin n}).

$u_{i}^{n,m*}$ is the specific thermal energy that is predicted at the end of iteration $m$. To get the energies used in the conductivity and gradients at the beginning of the next iteration, we interpolate between this value and the specific thermal energy at the beginning of this iteration, $u^{n,m-1}$, via
\begin{equation}
	\boldsymbol{u}^{n,m} = \mu^m \,\boldsymbol{u}^{n,m*} + (1-\mu^m)\, \boldsymbol{u}^{n,m-1} \, ,
\end{equation}
where $\boldsymbol{u}$ is a vector of thermal energies of all cells being considered in the calculation. A good choice of the interpolation parameter, $\mu^m$, can speed up the rate at which the non-linear solver converges. We implemented two different interpolation methods. First, similar to \citet{2016Kannan+}, we employ a modified version of the unstable manifold corrector scheme described in \citet{2010Smedt+} in which the correction vector (i.e. the vector of changes in thermal energy across the timestep) is adaptively over- or underrelaxed depending on the relative direction of successive correction vectors. In the second interpolation method, we take $\mu^m = 1$ but reduce this when necessary to ensure that the relative change of the thermal energy of all cells from its value at the beginning of the iteration is not more than $10\%$. We found the second procedure to be more robust, while the first method sometimes took much longer to converge, particularly when using local-timestepping. 

We use two criteria to determine if the solution is converged, both of which are checked at every iteration. 
The first criterion we check is
\begin{equation}
\label{eq: convg 1}
	\frac{|\boldsymbol{u}^{n,m} - \boldsymbol{u}^{n,m-1}|}{|\boldsymbol{u}^{n,0}|} < 10^{-6} \, .
\end{equation}
For the second convergence criterion we begin by estimating the solution at the end of the iteration using a forward-Euler scheme for both the tangential and normal components. The first part of this is identical to the standard treatment of the tangential fluxes (see equation~\ref{eq nonlin t}), so only the normal component requires calculating at each iteration. 

The estimate of the solution at the end of the iteration, $u_{i}^{\hat{n},m}$, is then
\begin{equation}
	u_{i}^{\hat{n},m} =  u_i^{\tilde{n},m} + \frac{\Delta t}{m_i^{n+1} c_{\rm v}}\sum_{j} \bigg[ \chi_{ij}^{\tilde{n},m} \, (\boldsymbol{b}_{ij}\boldsymbol{\cdot}\boldsymbol{\nabla} u_{ij,{\rm N}}^{\tilde{n},m})\,(\boldsymbol{b}_{ij}\boldsymbol{\cdot}\boldsymbol{A}_{ij})\bigg] \, , 
\end{equation}
where $u_{i}^{\tilde{n},m}$ is given by equation~(\ref{eq nonlin t}). We then calculate the difference between these predictions, $\boldsymbol{u}^{\hat{n},m}$, and the internal energies at the beginning of the iteration  $\boldsymbol{u}^{n,m}$,
\begin{equation}
	\Delta \boldsymbol{u}^{n,m} = \boldsymbol{u}^{n,m} - \boldsymbol{u}^{\hat{n},m} \, ,
\end{equation}
and the second convergence criterion is then
\begin{equation}
	\frac{|\Delta \boldsymbol{u}^{n,m}|}{|\Delta \boldsymbol{u}^{n,0}|} < 10^{-3} \, .
\end{equation}
If either of these two conditions are satisfied then we stop the iteration and accept the internal energies at the end of the final iteration.

We find that convergence typically occurs within the first $\sim 10$ iterations but can sometimes take many more, particularly when using local-timestepping (as described below) at synchronisation points where a significant fraction of gas cells are active. We should note, however, that these procedures are not strictly guaranteed to converge to the solution of the equations, in particular, due to the flux and gradient limiters that are used. Additionally, the first criterion essentially measures whether the iteration has stopped and does not make any statement about how close the solution is to solving the equations.

\begin{figure*}
    \centering
    \includegraphics[width=0.98\textwidth]{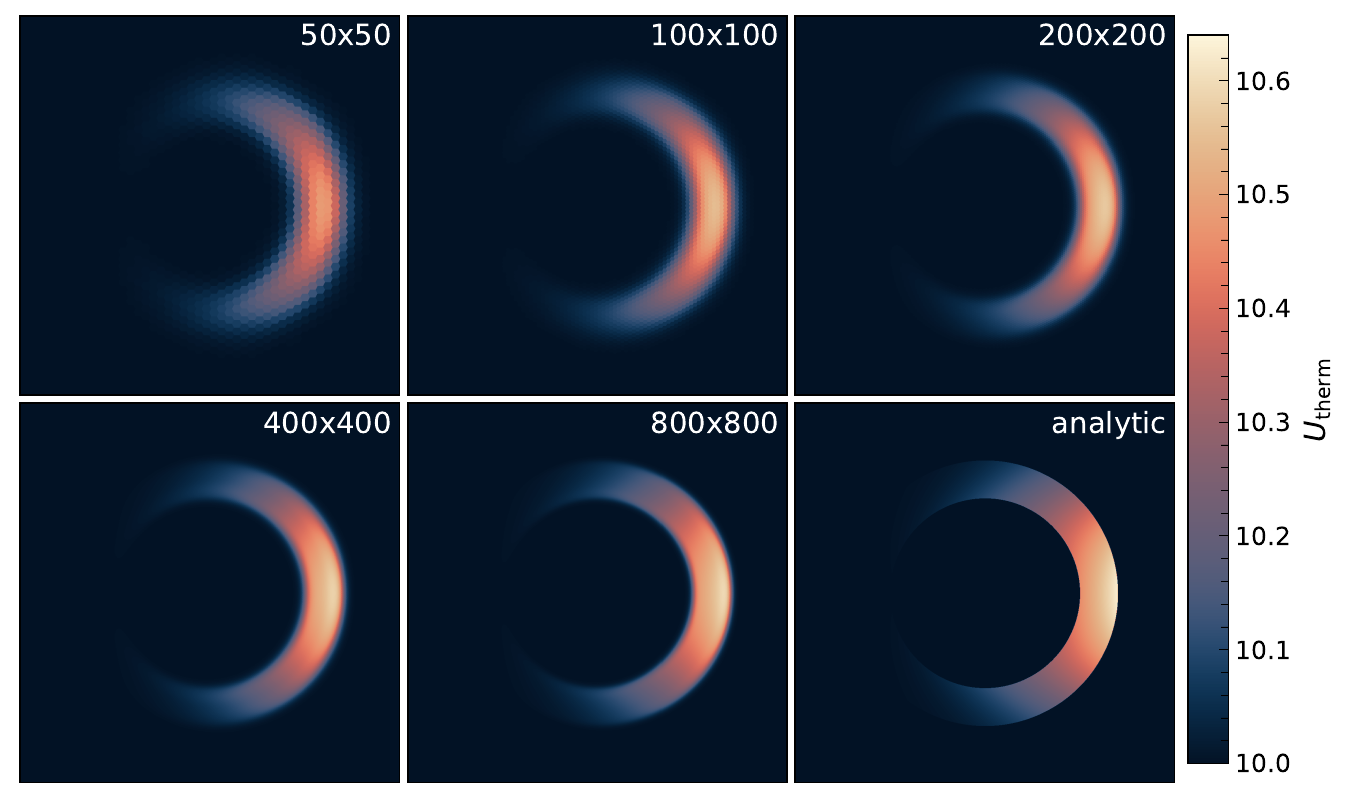}
    \caption[Ring vis]
    {Specific internal energy slices at early times ($t=10$) for the problem of anisotropic conduction of a hot patch of gas around a circular magnetic field in 2D. Each panel shows the entire computational domain for five simulations with different resolutions (indicated in the top right) which use local timestepping (see text for explanation). In the lower-right panel we show the analytic solution for comparison. With increasing resolution the diffusion perpendicular to the magnetic field decreases and the numerical solution is seen to converge on the analytic one.}
    \label{fig: Ring viz}
\end{figure*}

\begin{figure}
    \centering
    \includegraphics[width=0.49\textwidth]{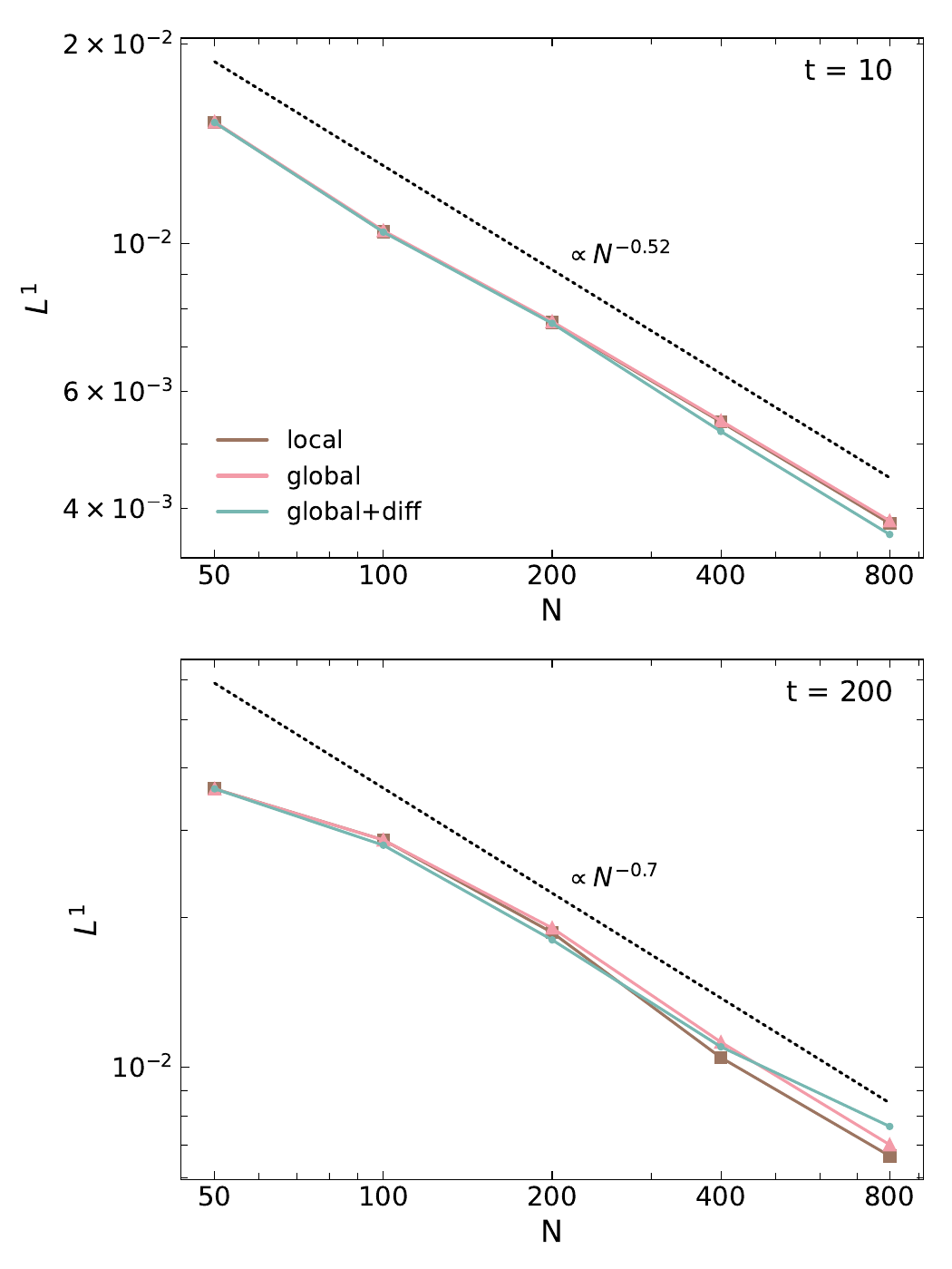}
    \caption[L1 norm ring]
    {$L^1$ norm for the problem of anisotropic conduction of a hot patch of gas around a circular magnetic field in 2D (as visualised in Fig.~\ref{fig: Ring viz}). The top panel shows the $L^1$ norm at early times ($t=10$), and the bottom panel shows it at late times ($t = 200$). The different coloured lines show the results for different timestepping schemes: `local' and `global' correspond to simulations that use local and global timestepping schemes, respectively. `global+diff' corresponds to simulations that use global timestepping with the addition of the timestep criterion required by explicit schemes for stability (equation~\ref{eq: tstp}). In general, all configurations show very similar errors and convergence rates, both at early times ($L^{1} \propto N^{-0.52}$) and at late times ($L^{1} \propto N^{-0.7}$).}
    \label{fig: L1 norm ring}
\end{figure}

\subsubsection{Local timestepping}
\label{subsec: local timestepping}
The scheme presented thus far solves the conduction problem for all gas cells in the simulation at each timestep. Since, when using local-timestepping, the mesh in {\small AREPO} is only guaranteed to be complete for active cells, the scheme, as currently presented, can only be used on global timesteps. This makes it unsuitable for use in complex problems with deep timestep hierarchies (such as those typically found in simulations of galaxy clusters). We, therefore, use a method similar to \citet{2016Pakmor+} to adapt the scheme to work with local timestepping.

At each timestep we compute the specific thermal energy of each cell involved in the calculation and use this to solve the conduction problem for each active interface\footnote{An active interface refers to a Voronoi face where at least one of the two neighbouring cells is active.} as described in the previous sections. We then update the thermal energy of the cells due to these conduction fluxes.
Note that the mesh at each timestep is guaranteed to be complete for active cells so all information required to calculate the gradients and quantities at the corners will be available. This means that we only have to solve the conduction problem for all active cells and for a layer of inactive cells that share a face with an active cell. 

For the `semi-implicit, non-linear solver' we must additionally define the particle set over which we calculate the convergence criteria (see Section~\ref{subsub: non linear solver}). We choose this set to be all cells that have at least one active interface, (i.e. all cells that can potentially have their internal energies changed by the thermal conduction in this timestep). Note that this is the same particle set that is used in the matrix iteration stopping criterion in the implicit step (used in both the linear and non-linear solvers) as the matrix will include the energies of all cells that are involved in the flux calculation (i.e. those that have at least one shared interface).

The scheme presented in this work uses a fundamentally different method to that of the anisotropic thermal conduction solver described in \citet{2016Kannan+} which is also implemented in {\small AREPO}. In addition, extending our scheme to work with local timesteps using the procedure described above is relatively straightforward. Building the mesh at each sync-point for the active particle set requires inserting the direct neighbours of all active cells into the mesh which, as discussed above, is sufficient for us to compute the fluxes over all active interfaces. The scheme presented in \citet{2016Kannan+}, however, requires information about a second layer of cells around active cells, which are not guaranteed to be part of the partial mesh. This means that the scheme can only be used on timesteps where all particles are active, making it unsuitable for use in investigating complex problems (such as those found in studies of structure formation) with deep timestep hierarchies.

\subsection{Anisotropic thermal conduction in comoving coordinates}
\label{subsec: comoving}
Here we briefly discuss the changes that need to be made to adapt the procedure described in the previous sections for use in cosmological simulations. In cosmological simulations it is convenient to introduce spatial coordinates that are comoving with the cosmological expansion and a set of comoving variables (which we indicate with a subscript `$\rmn{c}$'). Those relevant to the transformation of the conduction equation are defined via
\begin{align}
\label{eq: subs}
	&\boldsymbol{r} = a \boldsymbol{x}\, , \quad \boldsymbol{u} = \boldsymbol{\varv} - \dot{a}\boldsymbol{x} \, , \nonumber \\
    &\rho = \rho_\rmn{c} \, a^{-3} \, , \quad P = P_\rmn{c} \, a^{-3} \, \quad \chi = \chi_\rmn{c} \, a^{-1}\, , 
\end{align}
where $a$ is the scale factor, $\boldsymbol{r}$ are physical coordinates, $\boldsymbol{x}$ are comoving coordinates, the physical velocity is $\boldsymbol{\varv} = \boldsymbol{\dot{r}}$, and the peculiar velocity is $\boldsymbol{u} = a \, \boldsymbol{\dot{x}}$. Note that, under this coordinate transformation, the specific internal energy is unchanged. 

The derivatives in comoving coordinates transform according to 
\begin{equation}
	\frac{\partial}{\partial t}\bigg|_{\,\boldsymbol{r}}  = \frac{\partial}{\partial t}\bigg|_{\,\boldsymbol{x}} - H\,\boldsymbol{x} \boldsymbol{\cdot} \boldsymbol{\nabla}_{\boldsymbol{x}}\, , \qquad \boldsymbol{\nabla}_{\boldsymbol{r}} = \frac{1}{a}\boldsymbol{\nabla}_{\boldsymbol{x}}\, ,
\end{equation}
where $H = H(a) = \dot{a}/a$ is the Hubble rate.

The part of the energy equation relevant for anisotropic thermal conduction in comoving coordinates is then 
\begin{equation}
\label{eq: atc comoving}
	\frac{\partial u}{\partial t}\bigg|_{\,\boldsymbol{x}} = \frac{1}{\rho_\rmn{c} c_{\rm v}}\, \boldsymbol{\nabla}_{\boldsymbol{x}} \boldsymbol{\cdot} \bigg[ \chi_\rmn{c} \, \boldsymbol{b}(\boldsymbol{b}\boldsymbol{\cdot}\boldsymbol{\nabla}_{\boldsymbol{x}}\, u)\bigg] \, .
\end{equation}
Note that this equation does not correspond to an exact transform of equation~(\ref{eq: atc}) but, rather, comes about by transforming the energy equation~(\ref{eq: energy}) and isolating the term relevant for anisotropic thermal conduction. 
The discretised form of this equation is then
\begin{equation}
\label{eq: spatial disc com}
	\frac{\partial u_i}{\partial t}\bigg|_{\,\boldsymbol{x}} = \frac{1}{m_i c_{\rm v}}\sum_{j} \bigg[ \chi_{ij, {\rm c}} \, (\boldsymbol{b}_{ij}\boldsymbol{\cdot}\boldsymbol{\nabla}_{\boldsymbol{x}}\, u_{ij}) \, (\boldsymbol{b}_{ij}\boldsymbol{\cdot}\boldsymbol{A}_{ij, {\rm c}})\bigg] \, .
\end{equation}

\section{Results: Numerical tests}
\label{sec: Tests}
In this section we present results from a number of numerical test problems which we use to assess the accuracy and stability of the anisotropic thermal conduction scheme.

\subsection{Anisotropic conduction around a ring in 2D}
\label{subsec: Ring}

We test the 2D diffusion of a hot patch of gas in a circular magnetic field, following the setup of \citet{2005ParrishStone} and \citet{2007SharmaHammett}. We use this setup to test and quantify the convergence of the anisotropic solver with respect to the analytical solution when applied to a multidimensional problem at fixed conductivity and a constant background. We, therefore, disable hydrodynamics in this test and keep the mesh fixed.

We use a regular, approximately hexagonal mesh, which we create by taking a uniform Cartesian mesh offsetting the points in every other column by $0.49$ times the cell size. We initialise the specific thermal energy within a domain of $[-1,1]^2$ to
\begin{equation}
  u(\boldsymbol{x}) =
    \begin{cases}
      12 & \text{if $0.5 < r < 0.7$ and $|\phi| < \pi / 12$} \, ,\\
      10 & \text{otherwise} \, ,
    \end{cases} 
\end{equation}
where $r = \sqrt{x^2 + y^2}$ is the distance from the centre of the box and $\phi = \tan^{-1}( y/x)$ is the angle to the $x$-axis. The magnetic field is initialised via
\begin{align}
    &B_{x}(\boldsymbol{x}) = -\frac{y}{r} \, ,\nonumber\\
    &B_{y}(\boldsymbol{x}) = +\frac{x}{r}\, .
\end{align}
We set the parallel diffusivity, $\kappa = \chi / (c_{\rm v} \rho)$, to $0.01$. There is no perpendicular diffusivity so the energy should stay within the ring.

At early times, the resulting behaviour can be considered as a 1D diffusion problem of a double step function, which has an analytic solution given by
\begin{equation}
u(\boldsymbol{x}) = 10 + {\rm erfc}\bigg[\bigg(\phi + \frac{\pi}{12}\bigg) \frac{r}{D}\bigg] - {\rm erfc}\bigg[\bigg(\phi - \frac{\pi}{12}\bigg) \frac{r}{D}\bigg]\, ,
\end{equation}
where $D = \sqrt{4\kappa t}$ for $0.5 < r < 0.7$ and $u(\boldsymbol{x}) = 10$ elsewhere. At late times, the energy should be uniformly distributed in the ring, i.e.~$u = 10 + 1/6$ for $0.5 < r < 0.7$ and $u = 10$ elsewhere.

In Section~\ref{subsub: non linear solver} we introduced the non-linear solver as a way to take the temperature dependence of the conduction coefficient into account. In this test the conduction coefficient is fixed and therefore insensitive to the non-linear solver. We therefore defer discussion of the non-linear solver to the following section and, for this test, consider only the semi-implicit, linear solver described in Section~\ref{subsubsec: linear solver}.

In the present section we compare simulations that use local timesteps with those that use global timesteps and additionally consider simulations where we impose the timestep criterion required by explicit schemes for stability (equation~\ref{eq: tstp}). We refer to these three types of simulation as `local', `global' and `global+diff', respectively.

In the simulations that use local timesteps we impose a timestep hierarchy, as this does not arise naturally in the problem. We leave the timesteps of cells in the lower-left quadrant unaltered and reduce that of the cells in the upper-left and lower-right quadrant by a factor of $2$, and that of the cells in the upper-right quadrant by a factor of $4$.

Figure~\ref{fig: Ring viz} shows the specific internal energy distribution at early times ($t=10$) for five simulations with different resolutions, and in the bottom-right we show the analytic prediction. The simulations in Fig.~\ref{fig: Ring viz} were all run with local timestepping (as described above). Reassuringly, it is evident that there are no unphysical features at the boundaries between regions with different timesteps. All simulations reproduce the general behaviour predicted by the analytic solution. With increasing resolution, however, there is less perpendicular diffusion and the numerical solution becomes a better match to the analytic one.

\begin{figure*}
    \centering
    \includegraphics[width=\textwidth]{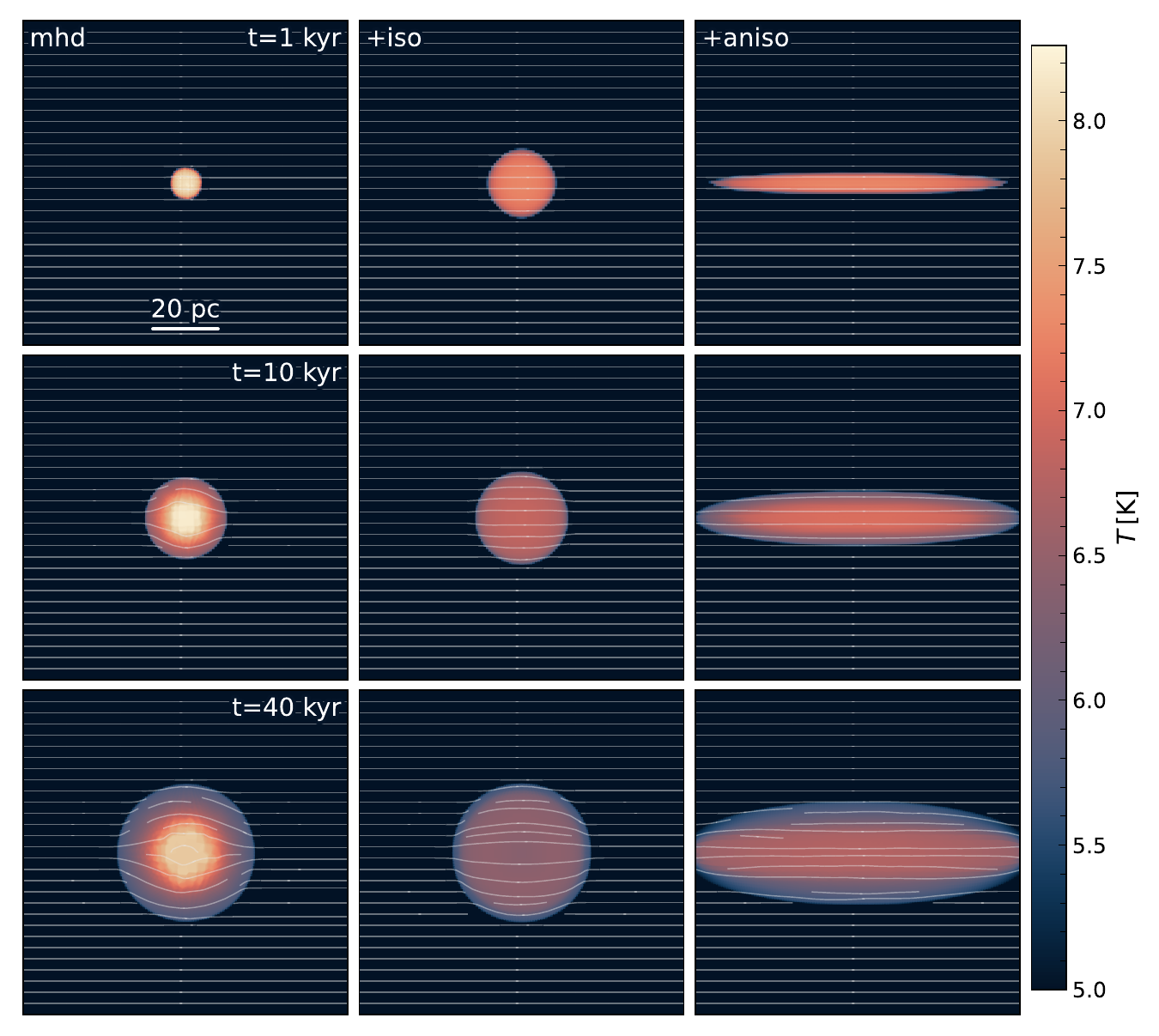}
    \caption[Sedov viz]
    {Thin temperature projections of size $100\times100$~pc for the point explosion test. Each column, from left to right, shows results for the pure MHD case, that with isotropic conduction and that with ansiotropic conduction. Each row depicts the state of the system at a different time which is indicated in the top-right corner of each panel in the first column. The grey lines show the magnetic field configuration. Thermal conduction can significantly affect the outcome of a point explosion, particularly when conduction occurs anisotropically along the magnetic field.}
    \label{fig: Sedov viz}
\end{figure*}

We now examine these simulations more quantitatively. Figure~\ref{fig: L1 norm ring} shows the $L^1$ norm\footnote{The sum over all cells of the absolute value of the difference between the simulation result and analytic prediction.} as a function of resolution of the simulation at early and late times, for simulations that use local timestepping and for runs that use global timestepping, with and without the diffusion timestep constraint being imposed. In general, all three configurations show very similar errors and convergence rates, at both early times ($L^{1} \propto N^{-0.52}$) and late times ($L^{1} \propto N^{-0.7}$).
The faster convergence rates at late times arise due to the fact that, by this point, the solution is primarily sensitive to errors in the conduction perpendicular to the magnetic field whereas any errors in the parallel conduction speed (which the solution will be sensitive to at early times) have been washed out. 
We note that these convergence rates are clearly worse than first order but are comparable to other implementations of anisotropic transport solvers \citep{2005ParrishStone, 2007SharmaHammett, 2016Kannan+, 2016Pakmor+}.  

The behaviour of the errors in the runs with local and global timesteps are very similar. At late times, however, the run with local timesteps has slightly smaller errors at higher resolution. This may arise because we impose smaller timesteps than required in the local-timestepping simulations. The runs with the additional timestep constraint show slightly smaller errors at early times and higher resolution in comparison to the other schemes. At late times and high resolution, however, the error is slightly worse. 
 
\subsection{Point explosion in a uniform medium}
\label{subsec: sedov}

To assess the accuracy of the coupling between the ideal MHD and thermal conduction solvers we carry out 3D simulations of a point explosion in a uniform background where the conductivity is self-consistently determined by the temperature. We perform three different types of simulation; one where there is no thermal conduction (i.e. ideal MHD), one with the addition of isotropic conduction and one with anisotropic conduction. We refer to these type of simulations as `MHD', `+iso' and `+aniso', respectively.

In all simulations, the mesh is initially Cartesian but undergoes regularisation during the course of the simulation \citep[see][]{2012Vogelsberger+}. We carry out simulations at three different spatial resolutions: $64^3$, $128^3$ and $256^3$. For the ideal MHD and isotropic conduction simulations, the computational domain has side length $100$~pc. As the conduction front in the setup with anisotropic conduction moves much faster than in the isotropic case we use a computational domain with side length $200$~pc for the two lower resolution simulations with anisotropic conduction. This ensures that the conduction front does not leave the box during the simulation. For the highest resolution simulation with anisotropic conduction, increasing the boxsize was too expensive and so we ran it in a box of side length $100$~pc and stopped the simulation before the conduction front left the domain. For consistency we will continue to refer to the lower resolution simulations with a box side length of $200$~pc as $64^3$ and $128^3$, despite the fact that, when doubling the domain length, we also double the number of cells along each edge. This ensures that, across all simulations of a given resolution (i.e. $64^3$, $128^3$ or $256^3$), the spatial resolution remains the same.

The background medium has constant density $n_0 = 1 \, {\rm cm^{-3}}$ and temperature $T_0 = 10^4 \, {\rm K}$. The background magnetic field has strength $1\, \mu{\rm G}$ and points in the positive $x$-direction, which corresponds to an initial plasma beta of $\sim 35$. In all simulations we inject $E_0 = 3.33\times10^{50} \, {\rm erg}$ into the central region. In the $64^3$ resolution runs we inject into the central $8$ cells. We keep the volume of the injection region the same in the higher resolution simulations ($128^3$ and $256^3$), and the number of injection cells increases accordingly to $64$ and $512$, respectively. This means that the total mass into which the energy is injected is constant across all simulations (and thus the peak initial temperature is also the same). 

For simulations with conduction, the conductivity is assumed to take the Spitzer value without any form of suppression (i.e. $\chi$ is as in equation~\ref{eq: spitzer}). Unless specified, the simulations were run with the linear thermal conduction solver. We identify those that were run with the non-linear solver explicitly. 

In order to capture the very early evolution of the shock and conduction fronts, we employ maximum timesteps that are log-spaced in time. We, additionally, carried out analogous simulations with local timesteps where we relaxed this maximum timestep criterion and confirmed that the results of these simulations are consistent with the late time behaviour described in the rest of this section.

Figure~\ref{fig: Sedov viz} shows thin temperature projections at three different times for the $128^3$ resolution simulations without conduction, with isotropic conduction and with anisotropic conduction. At early times, the conduction fronts in the runs with conduction (middle and right columns) advance faster than the shock front in the pure MHD run (left column). In the simulation with anisotropic conduction there is essentially no diffusion perpendicular to the magnetic field and the parallel expansion occurs noticeably faster than the radial expansion seen in the simulation with isotropic conduction. The perpendicular expansion in the anisotropic conduction case is also slower than the classical adiabatic solution, as was also found in \citet{2016Kannan+} and \citet{2016DuboisCommercon}. 

At later times, the pure MHD and isotropic conduction simulations have similar radial extents although the MHD simulation retains a strong radial temperature gradient within the shocked region, while the temperature gradients in the simulations with conduction are much flatter. 


We now examine the different behaviour observed in these simulations more quantitatively. Analytic solutions to the problem of point explosion in a uniform medium exist for the purely adiabatic case and the case of pure conduction. For an adiabatic point explosion, the radius of the shock front expands according to
\begin{equation}
\label{eq: r sedov}
    r_{\rm s}(t) = \xi(n)\bigg(\frac{E_0\,t^2}{\rho_0}\bigg)^{1/(n + 2)}\, ,
\end{equation}
where $n$ is the dimension of the problem and $\xi(n)$ is a constant of order unity which also depends on the adiabatic index \citep{1959Sedov}. In three dimensions, with $\gamma = 5/3$, $\xi(3)$ is approximately $1.15$.

For the case of a point explosion with heat conduction where the conductivity has a power law dependence on temperature, $\chi = \chi_0\, T^\alpha$, the radius of the conduction front evolves according to 
\begin{equation}
\label{eq: r cond}
    r_{\rm c}(t) = \Bigg[\bigg(\frac{2(n\alpha + 2)}{\alpha}\frac{\chi_0}{c_{\rm v} \rho_0}\bigg)\bigg(\frac{2}{S_n} \frac{1}{\mathcal{B}(\frac{n}{2},\frac{1}{\alpha} + 1)}\frac{E_0}{\rho_0 c_{\rm v}}\bigg)^\alpha t\Bigg]^{1/(n\alpha+2)}\, ,
\end{equation}
where $S_n = 1$, $2\pi$, $4\pi$ for dimensions $n=1$, $2$, $3$, and $\mathcal{B}(x,y)$ is the beta function \citep[see e.g.][]{1967ZeldovichRaizer, 1996Barrenblatt}.

\subsubsection{Point explosion with isotropic conduction}
\label{subsub: sedov iso}
\begin{figure}
    \centering
    \includegraphics[width=0.49\textwidth]{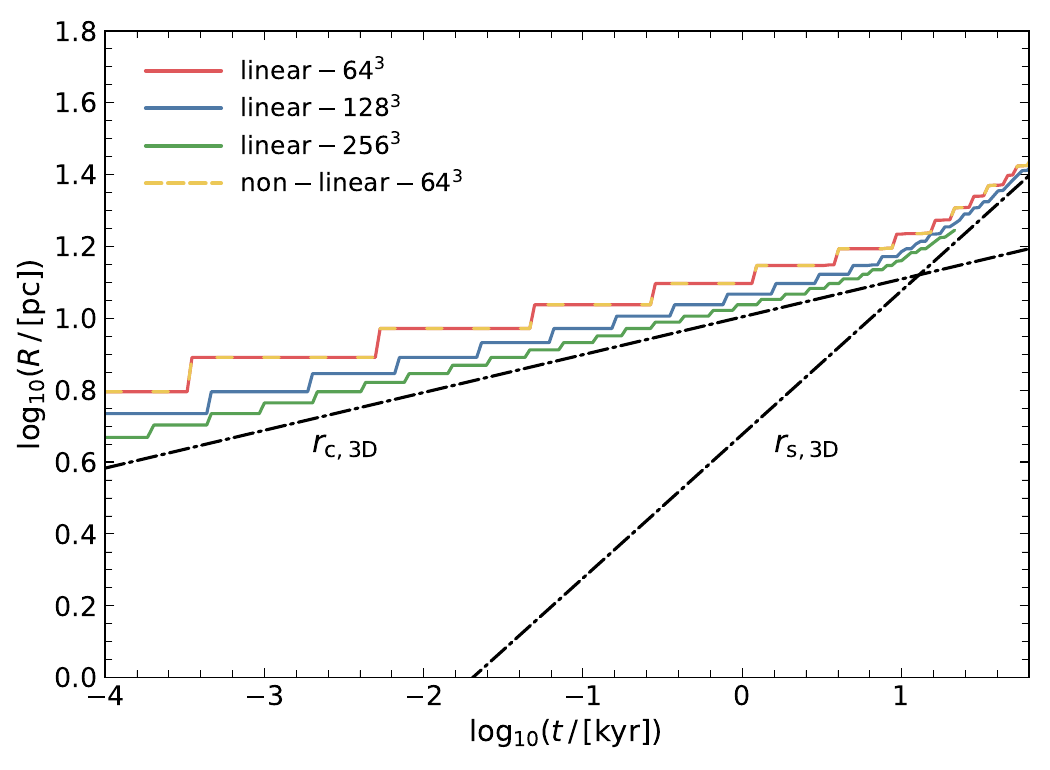}
    \caption[Sedov iso]
    {Evolution of the radius of the conduction/shock front in the 3D point explosion test with isotropic conduction (as visualised in the middle column of Fig.~\ref{fig: Sedov viz}). Solid, coloured lines show results from simulations run with the linear conduction solver while the dashed, yellow line corresponds to a simulation run with the non-linear solver (see figure legend). The black lines indicate the analytic expectation for the radial evolution of the shock/conduction front in the cases of pure hydrodynamics ($r_{\rm s,3D}$, equation~\ref{eq: r iso ad}) and pure conduction ($r_{\rm c,3D}$, equation~\ref{eq: r iso cond}). At early times conduction is fastest but at $\sim13$~kyr, the shock front catches up and overtakes. With increasing resolution the results converge towards the analytic expectation.}
    \label{fig: Sedov iso}
\end{figure}

We begin by examining the simulations that were run with isotropic conduction. Equation~(\ref{eq: r sedov}) predicts that the 3D shock radius in our simulations should expand according to
\begin{equation}
\label{eq: r iso ad}
    r_{\rm s, 3D}(t) = 4.75 \, \bigg(\frac{t}{{\rm kyr}}\bigg)^{\frac{2}{5}} \, {\rm pc} \, .
\end{equation} 
From equation~(\ref{eq: r cond}) the 3D conduction front, assuming Spitzer conductivity ($\alpha=5/2$), is at radius
\begin{equation}
\label{eq: r iso cond}
    r_{\rm c, 3D}(t) = 10.1 \, \bigg(\frac{t}{{\rm kyr}}\bigg)^{\frac{2}{19}} \, {\rm pc} \, .
\end{equation}

In Fig.~\ref{fig: Sedov iso} we show the time evolution of the radius of the shock/conduction front for the three different resolution simulations. Additionally, for the lowest resolution, we show the results from a simulation run using the non-linear conduction solver described in Section~\ref{subsub: non linear solver}.  We estimate the radius of the shock/conduction front by first calculating a spatial temperature profile of the gas along the $x$-axis\footnote{The results discussed in the remainder of the section are independent of the direction of the axis along which the temperature profile is measured.}, passing through the centre of the injection region. We then define the conduction/shock front radius to be the distance from the centre to the furthest point where the temperature is above $1\%$\footnote{We also tested threshold values of $0.1\%$ and $10\%$ and found that the location of the shock/conduction front radius is sufficiently well resolved in all simulations that the results presented in this section and the one that follows are insensitive to the exact choice of this threshold value.} of the background temperature, $T_0$.

At early times, conduction dominates and the radius of the fronts advances $\propto t^{2/19}$, in accordance with analytical expectations. Equating equations~(\ref{eq: r iso ad}) and (\ref{eq: r iso cond}) gives a transition time of $\sim13$~kyr at which hydrodynamical process are expected to become dominant. Indeed, this is very close to the time at which we see a distinct change in the evolution of the shock/conduction front in the simulations, which then begins to follow the shock solution. With increasing resolution and at a fixed time, the radius of the measured front is generally smaller, and converges towards the analytic expectation. The behaviour of the shock/conduction front in the simulation run with the non-linear solver (dashed yellow line) almost exactly reproduces that in the analogous simulation run with the linear solver (solid red line). 
  
\subsubsection{Point explosion with anisotropic conduction}

\begin{figure}
    \centering
    \includegraphics[width=0.49\textwidth]{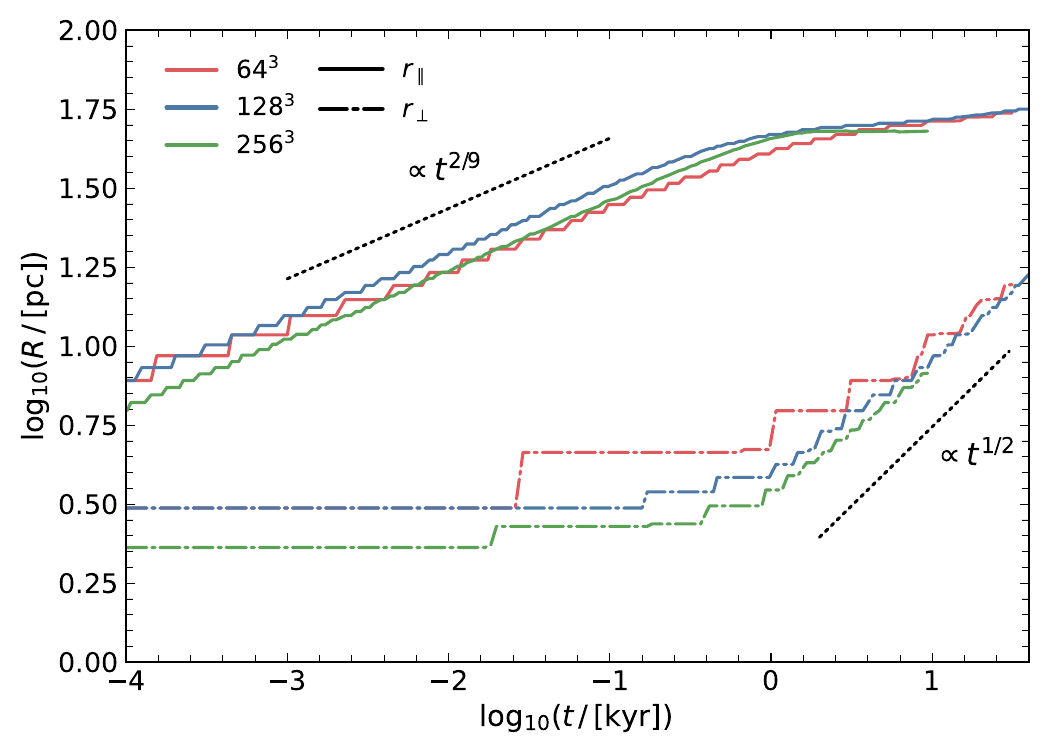}
    \caption[Sedov aniso]
    {Evolution of the radius of the conduction/shock front in the 3D point explosion test with anisotropic conduction (as visualised in the right column of Fig.~\ref{fig: Sedov viz}). Solid/dot-dash, coloured lines show the radius as measured parallel/perpendicular to the initial magnetic field direction (see figure legend). The black dotted lines show the analytic scalings for the radial evolution of a cylindrical blast wave ($r\propto t^{1/2}$, see equation~\ref{eq: r sedov}) and a 1D conduction front ($r\propto t^{2/9}$, see equation~\ref{eq: r cond}). The expansion of the conduction front along the magnetic field follows the expected scalings for a 1D point explosion with pure conduction. Perpendicular to the initial magnetic field, the shock front expands slower than the conduction front and follows the expected scalings for a cylindrical blast wave.}
    \label{fig: Sedov aniso}
\end{figure}

\begin{figure*}
    \centering
    \includegraphics[width=\textwidth]{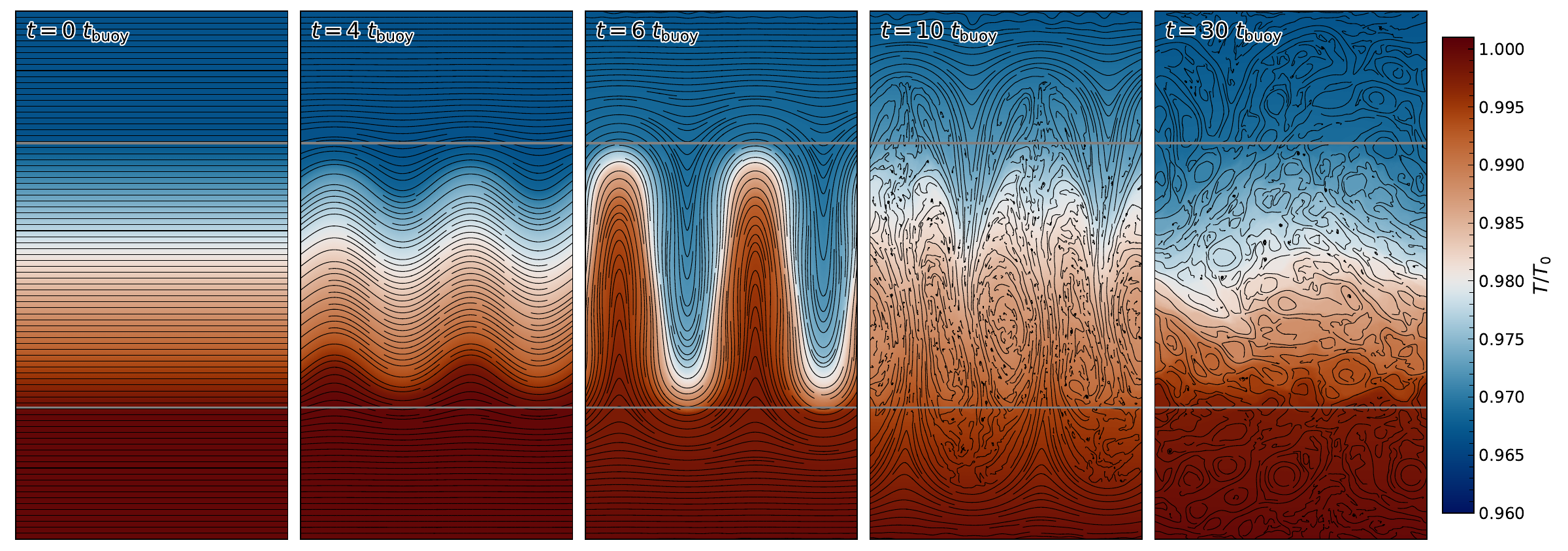}
    \caption[MTI temp]
    {Temperature slices with magnetic field lines overlaid in black that show the time evolution of the MTI for the simulation where the unstable region has resolution $256\times256$. The slices show the entire simulation domain ($L\times2L$), and the grey horizontal lines indicate the boundaries between the unstable and neutral regions. The temperatures are normalised to that at the base of the unstable region. The initial perturbation grows linearly in the first three panels until there is a large vertical component of the magnetic field. This configuration is, however, non-linearly unstable with the end result being a state of sustained turbulence.}
    \label{fig: mti temp}
\end{figure*}

When heat conduction is anisotropic, the problem of point explosion in a uniform medium essentially becomes a 1D problem along the direction of the magnetic field. Perpendicular to the magnetic field lines, the evolution of the system is determined by MHD processes and the early rapid expansion of the conduction front along the $x$-axis, as seen in Fig.~\ref{fig: Sedov viz}, means that this approximately resembles a 2D cylindrical blast wave. 

Equations~(\ref{eq: r sedov}) and (\ref{eq: r cond}) show that the radii of the conduction and shock fronts in this scenario should scale as $r_{\rm c,1D} \propto t^{2/9}$ and $r_{\rm s, 2D} \propto t^{1/2}$, respectively. In contrast to the isotropic case, the constants $E_0$ and $\rho_0$ in equations~(\ref{eq: r sedov}) and (\ref{eq: r cond}) are not well defined because the anisotropic case does not decouple into two independent problems along and perpendicular to the magnetic field. We, therefore, just compare the scaling behaviour of the radius of the 2D shock and 1D conduction front, rather than exact values.

In Fig.~\ref{fig: Sedov aniso} we show the radii of the conduction/shock front in the point explosion test with anisotropic conduction as a function of time, for the three different resolution simulations. We calculate the radii of the shock/conduction front parallel and perpendicular to the initial magnetic field direction and use the same method as was described in the previous section for the isotropic conduction simulations. The expansion of the conduction front along the magnetic field at early times is much faster than seen for the isotropic case. This is expected and is largely due to the geometry of the problem. The radius of the conduction front follows the analytic scaling well for the higher resolution simulations, but is slightly shallower for the lowest resolution run (red line). At late times, the radial expansion slows with respect to the analytic scaling as hydrodynamical processes that are driving the lateral expansion become important. This behaviour was also seen in \citet{2016Kannan+}.

The expansion of the shock front perpendicular to the magnetic field is initially slower than that of the conduction front along the magnetic field and follows the expected scaling well at all resolution levels, except perhaps at very early times for the lowest resolution run. Overall, the results in this section have validated the accuracy of the coupling between the hydrodynamics and thermal conduction solvers and have, additionally, demonstrated the accuracy of our treatment of the temperature dependence of the conductivity in the solver.

\subsection{The magneto-thermal instability}

For both positive and negative temperature gradients, high-$\beta$, weakly collisional, stratified plasmas are unstable to magnetically-mediated buoyancy instabilities when anisotropic conduction is rapid compared to the dynamical response of a plasma. Specifically, anisotropic heat conduction causes the slow magnetosonic wave to become buoyantly unstable to the MTI \citep{2000Balbus, 2001Balbus} when the temperature decreases with height and to the HBI \citep{2008Quataert} when the temperature increases with height. 

In regions where $\boldsymbol{g}\boldsymbol{\cdot}\boldsymbol{\nabla} T > 0$, any misalignment of magnetic-field lines and gravity (i.e. $\boldsymbol{g}$) can result in the development of the MTI. In the presence of a sustained temperature gradient, the MTI drives vigorous turbulence that is largely isotropic \citep[see e.g.][]{2005ParrishStone, 2007ParrishStone, Perrone2022a, Perrone2022b}. In this section, we present simulations which explore the non-linear behaviour of the MTI and compare our results to previous work \citep[such as][]{2005ParrishStone, 2008ParrishQuataert, 2011McCourt+, 2016Kannan+}.
 
\subsubsection{Initial conditions}

We follow the evolution of a 2D region of plasma initially in hydrostatic and thermal equilibrium.  The region of plasma we simulate consists of an MTI unstable region, where conduction is anisotropic, sandwiched between two buoyantly neutral regions, where the heat conduction is isotropic. Note, however, that we ignore these buffer regions in the quantitative analysis presented in this section.

The plasma is stratified in density and temperature and is subject to a uniform gravitational field in the vertical direction: $\boldsymbol{g} = -g_0 \, \boldsymbol{\hat{z}}$. 
In the horizontal direction, we apply periodic boundary conditions. At the upper and lower boundaries, we fix the temperature. This enforces a vertical temperature gradient across the box and prevents the instabilities from saturating by flattening this gradient. We extrapolate the pressure into the ghost cells at these boundaries so as to enforce hydrostatic equilibrium, while all other plasma quantities are treated using reflective boundary conditions.

The size of the simulated domain is $L \times 2L$ and the unstable region lies within $L/2 < z < 3L/2$. In this unstable region we construct an atmosphere where the temperature decreases linearly with height according to
\begin{align}
    &T(z) = T_0 \,\bigg(1-\frac{z - L/2}{3H_0}\bigg) \, , \\
    &\rho(z) = \rho_0 \,\bigg(1-\frac{z - L/2}{3H_0}\bigg)^2 \, ,\\ 
    &P(z) = P_0 \,\bigg(1-\frac{z - L/2}{3H_0}\bigg)^3 \, ,
\end{align}
where $H_0$ is a scale height. We set $L/H_0 = 0.1$ which means that this setup is `local' in the sense that the size of the simulation domain is much smaller than the pressure scale height. 

We initialise the buoyantly neutral layers as isothermal atmospheres where the temperatures are continuous across the boundaries of the unstable region. The pressure and density in these regions vary exponentially with height.  We set $g_0 = k_\rmn{B}T_0/ (H_0 \,\mu\,m_{\rm H})$ to ensure that the system is initially in hydrostatic equilibrium and additionally apply a uniform background magnetic field in the $x$-direction with magnitude such that the plasma beta at the base of the unstable region is $\beta_0 = 2\times10^{4}$. This means that magnetic tension is negligible on the scales of interest.

The physics of the buoyancy instabilities is independent of the conductivity in the limit that the thermal diffusion time across the spatial scales of interest is short compared to the dynamical time. We, therefore, assume a uniform fixed conductivity and ensure that the simulations are in this fast conduction limit by setting $\chi = 10\, \rho_0 \, c_{\rm v} \, \omega_{\rm buoy} \, L^2$, where we evaluate the characteristic frequency for the buoyancy instability, 
\begin{equation}
    \omega_{\rm buoy} = \bigg|g_0\frac{\partial \ln T}{\partial z} \bigg|^{1/2}\, ,
\end{equation}
at the base of the unstable region. This means that the thermal diffusion time across the unstable region is $\sim 0.1\,\omega_{\rm buoy}^{-1}$. As the conductivity is fixed, we only present simulations that use the linear solver in this section. 

The simulations are initialised with a regular mesh, similar to that described in Section~\ref{subsec: Ring}, which is then allowed to move over the course of the simulation. We, additionally, consider two different spatial resolutions, where the unstable region is resolved by $128^2$ or $256^2$ cells. The simulations are all run for $50$ buoyancy times ($t_{\rm buoy} = \omega_{\rm buoy}^{-1}$, where $\omega_{\rm buoy}$ is evaluated at the base of the unstable region). We apply single mode transverse perturbations to the velocity field with wave vector $\boldsymbol{k} = (4\pi/L) \, \boldsymbol{\hat{x}}$ and amplitude $10^{-4} c_{\rmn{s},0}$ where $c_{\rmn{s},0}$ is the sound speed evaluated at the base of the buoyantly unstable region.

In 2D, when the mesh is allowed to move, we found that not resolving the gradients and the magnetic field topology could lead to the introduction of noise in the simulations. In these simulations we, therefore, applied an extra criterion whereby the tangential fluxes across a face are set to zero if the signs of the normal components of the magnetic field estimates at the two corners of each face are different. This increases the stability of the scheme at the cost of being more diffusive. We find that this condition is not necessary in 3D (where the number of corners that contribute to the interface estimate is significantly higher), or when the resolution is high enough.
\begin{figure*}
    \centering
    \includegraphics[width=\textwidth]{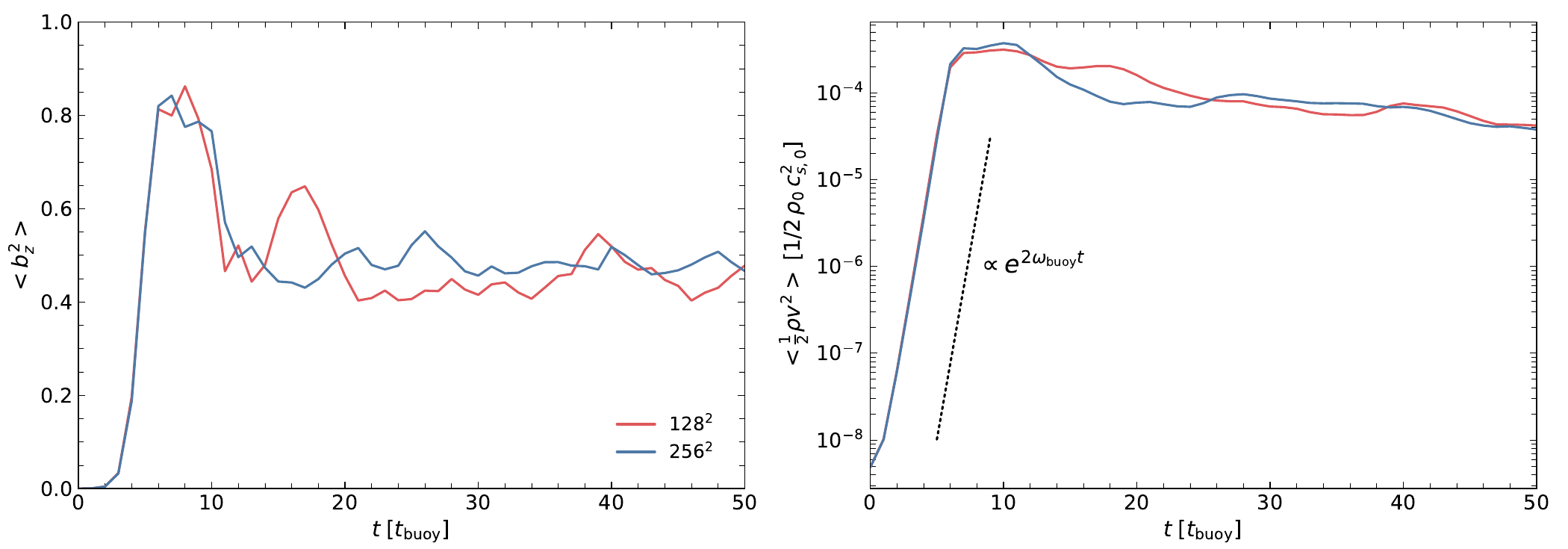}
    \caption[MTI bz energy]
    {The left-hand panel shows the time evolution of the magnetic field orientation in the two-dimensional MTI simulation visualised in Fig.~\ref{fig: mti temp}. Specifically, we show the volume average of the square of the vertical component of $\boldsymbol{b} = \boldsymbol{B}/|\boldsymbol{B}|$. A vertical magnetic field will have $b_z^2 = 1$, whereas $b_z^2=0.5$ corresponds to an isotropic field in 2D. The right-hand panel shows the time evolution of the volume averaged kinetic energy density. The different coloured lines correspond to simulations with different resolutions (see legend). These quantities were calculated by averaging over all cells within the buoyantly unstable region. The black dashed line shows the theoretical scaling of the growth rate. During the linear phase of the instability ($t \lesssim 6 \, t_{\rm buoy}$), the magnetic field is driven towards a vertical orientation and the kinetic energy is exponentially amplified. During the non-linear phase that follows, the instability saturates and the turbulence acts to isotropise the magnetic field.}
    \label{fig: mti bz energy}
\end{figure*}

\begin{figure*}
    \centering
    \includegraphics[width=\textwidth]{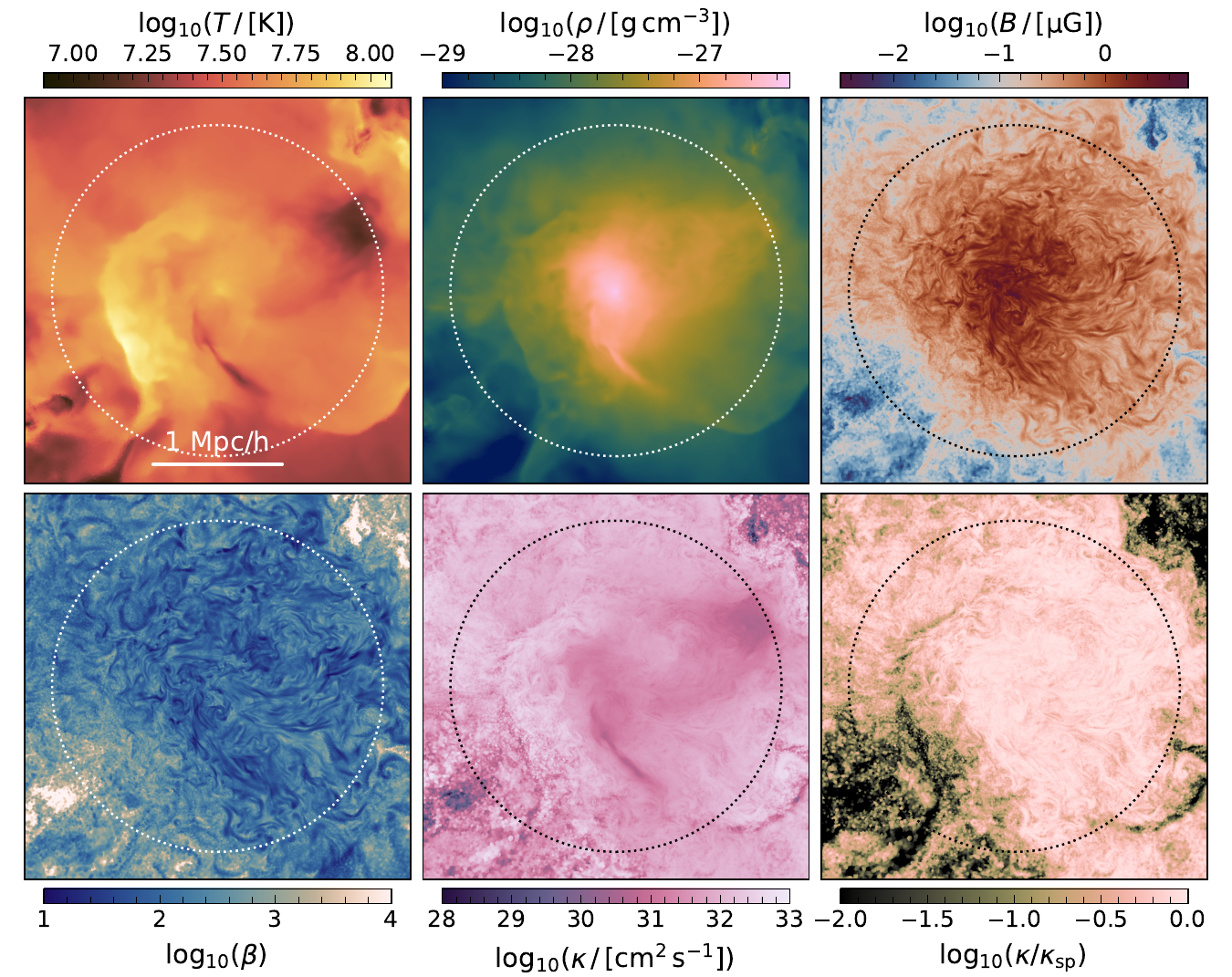}
    \caption[Cluster Viz]
    {Thin ($100\, h^{-1} \, {\rm kpc}$) projections of size $(3\,h^{-1}\,{\rm Mpc})^2$ at $z=0$ for the `+whistler' simulation. From left to right, the top row shows temperature, gas density and magnetic field strength. The bottom row shows the plasma beta, the thermal diffusivity, and the ratio of the (whistler suppressed) thermal diffusivity to the Spitzer value. The temperature projection is weighted by mass, while those of the magnetic field strength, density and thermal diffusivity are volume weighted. The projection of the plasma beta is calculated by dividing the volume weighted projection of the thermal pressure by that of the magnetic pressure; likewise, the projection of the ratio of the thermal diffusivity to the Spitzer value is calculated by dividing the volume weighted projection of the thermal diffusivity by that of the Spitzer value. Dashed circles correspond to $R_{\rm 200c}$. The magnetic field is highly turbulent and this, along with a much smoother temperature gradient leads to significant levels of suppression in the heat flux (see equation~\ref{eq: kappa interp}), particularly in the outskirts of the cluster and at shocks.}
    \label{fig: cluster viz}
\end{figure*}

\begin{figure*}
    \centering
    \includegraphics[width=\textwidth]{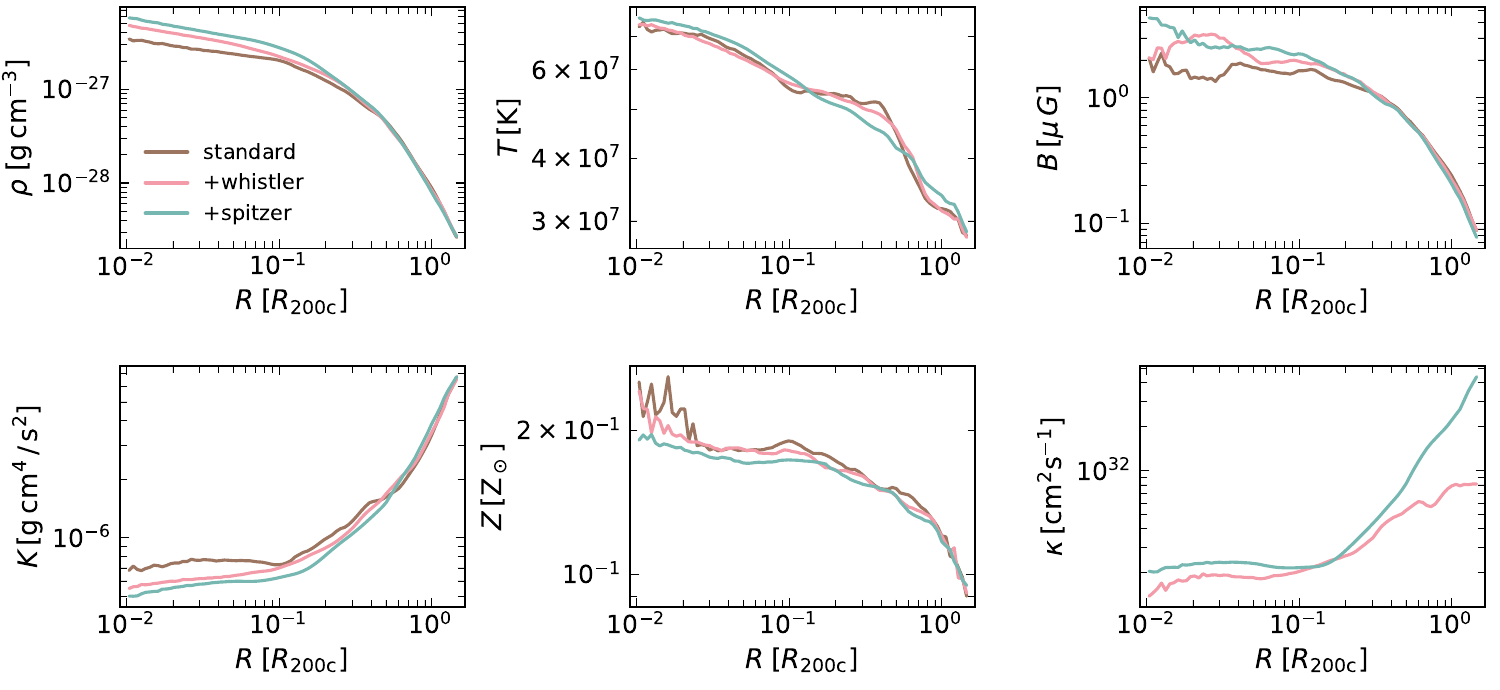}
    \caption[Cluster profiles]
    {Radial profiles at $z=0$ for the galaxy cluster zoom simulations. From left to right, the top row shows density, temperature, and magnetic field strength. The bottom row shows entropy (which we define to be $K=k_{\rm B} T / n_{\rm e}^{(\gamma - 1)}$, where $n_{\rm e}$ is the electron number density), metallicity, and thermal diffusivity, for the two simulations run with anisotropic thermal conduction. All profiles are in reasonably good agreement with each other, with some deviations seen in the central regions. The cluster simulated with full Spitzer conductivity (`+spitzer') shows larger differences than that simulated with whistler suppressed conduction (`+whistler') when compared to the simulation without conduction (`standard').}
    \label{fig: cluster profiles}
\end{figure*}

\subsubsection{Results}
Figure~\ref{fig: mti temp} shows temperature slices with magnetic field lines overlaid in black for the simulation where the unstable region has resolution $256\times256$. The slices show the entire simulation domain ($L\times2L$), including the upper and lower buoyantly neutral regions and the grey horizontal lines indicate the boundaries between these regions. In the first three panels, the instability is still in the linear growth phase. The perturbations grow exponentially and drive the development of a large vertical magnetic field component. 

The MTI growth rate goes to zero when the field lines become vertical. As we see from the final two panels of Fig.~\ref{fig: mti temp}, however, the MTI does not saturate quiescently when it reaches this linearly stable state but, instead, drives sustained turbulence. This arises due to the existence of zero-frequency modes of the dispersion relation, which correspond to horizontal perturbations that act to drive the plasma away from this equilibrium configuration \citep{Balbus2010, 2011McCourt+}. This state of sustained turbulence persists for the rest of the simulation. This qualitative behaviour is consistent with that observed in previous simulation work \citep[see e.g.][]{2005ParrishStone, 2007ParrishStone, 2011McCourt+, 2016Kannan+, 2021Berlok+}.

In the final two panels of Fig.~\ref{fig: mti temp} we see that the turbulent motions have length scales that are comparable to the size of the computational domain. As shown by \citet{2011McCourt+}, this prematurely stops the buoyant acceleration and implies that these `local' simulations are not able to accurately capture the saturated state of the MTI and under-predict the turbulent energies.

The left-hand panel of Fig.~\ref{fig: mti bz energy} shows the evolution of the $z$-component of the magnetic field and highlights the difference between the linear phase of the instability where the vertical component of the magnetic field is exponentially amplified and the saturated state where the magnetic field becomes close to isotropic. We also show the time evolution of the volume averaged kinetic energy density in the right-hand panel of Fig.~\ref{fig: mti bz energy}. In both simulations, the kinetic energy is exponentially amplified at early times but then saturates at $t \approx  6 \, t_{\rm buoy}$ as the magnetic field configuration approaches the linearly stable state. 

The linear growth rate of a mode of the MTI can be determined from the set of linearised equations that govern the system \citep[see e.g.][]{2008Quataert, 2011Kunz, 2021Berlok+}. Assuming locality in the vertical direction, the growth rate for the perturbation seeded in these simulations should be $\omega_{\rm buoy}$. The dashed line in the right-hand panel of Fig.~\ref{fig: mti bz energy} corresponds to this scaling and is in very good agreement with the behaviour observed in our simulations.

Both the lower and higher resolution simulations agree very well on the behaviour of the instability during the linear growth phase ($t \lesssim 6 \, t_{\rm buoy}$). At later times, the results differ quantitatively. These differences, however, are marginal and the qualitative behaviour is in good agreement. The evolution of the magnetic field and the kinetic energy densities, shown in Fig.~\ref{fig: mti bz energy}, are also very similar to those presented in \citet{2016Kannan+}, who carried out similar 2D simulations of the MTI, and to those found in the 3D simulations of \citet{2011McCourt+}.

\section{Cosmological simulations of galaxy clusters with anisotropic thermal conduction}
\label{sec: cluster}

To investigate the effects of anisotropic thermal conduction on structure formation and evolution it is important that the anisotropic thermal conduction solver is able to work efficiently and accurately in a fully cosmological context. In this section, we demonstrate that our code is capable of efficiently running high resolution cosmological zoom simulations of galaxy clusters. An in-depth analysis of the role of anisotropic thermal conduction in the evolution and properties of the cluster will be presented in a follow-up paper. Here we will largely focus on technical details and implementation choices.

In one of our simulations, we set the conductivity to the Spitzer value (equation~\ref{eq: spitzer}) and in the other we set it to Spitzer with whistler suppression (equation~\ref{eq: kappa interp}).

\subsection{Initial conditions and physical model}

In this section we present three different cosmological zoom simulations of one halo with mass $M_{\rm 200c} = 5.09\times 10^{14} \, h^{-1}\,{\rm M}_\odot$ and radius $R_{\rm 200c} = 1.30\, h^{-1} \, {\rm Mpc}$ at $z=0$.  Here, we define $M_{\rm 200c}$ and $R_{\rm 200c}$ such that the average density within a sphere of radius $R_{\rm 200c}$ is $200$ times the critical density of the universe and $M_{\rm 200c}$ is the mass enclosed by this radius. The simulations use a Planck-$2018$ cosmology \citep{2020Planck} where $\Omega_m = 0.315$, $\Omega_b = 0.049$, $\Omega_\Lambda = 0.684$, and the Hubble constant is $H_0 = 100\, h\, {\rm km \, s^{-1} \, Mpc^{-1}}$ with $h = 0.673$. 
The simulations presented in this section were carried out as part of the ongoing PICO-Clusters (Plasmas In COsmological Clusters) project, which will be presented in greater detail in forthcoming work. 

The particular halo studied here was selected for re-simulation from a parent dark matter only simulation with side of length $1 \,h^{-1}\, {\rm cGpc}$. The initial conditions for the zoom simulation were then created using a new code that will be described in Puchwein et al., in prep. 
We have made sure that the simulations presented here have no low resolution dark matter particles within $2\, R_{\rm 200c}$ at $z=0$.

In these simulations, the high resolution dark matter particles have mass $4.0\times10^7\, h^{-1} \, {\rm M_\odot}$ while the gas in the high resolution region has target mass resolution $7.4\times10^6\, h^{-1} \, {\rm M_\odot}$. The comoving softening length of the high resolution dark matter is set to $3.25\,h^{-1} \, {\rm ckpc}$ with a maximum physical softening length of $1.627\,h^{-1}\,{\rm kpc}$. The gas softening is treated adaptively and scales with the radius of the cell, with a minimum comoving value of $0.41\,h^{-1} \, {\rm ckpc}$.

This resolution is about $3$ times better than that used in the full-physics MillenniumTNG box \citep{2023Pakmor+}, and is slightly better than that used in the TNG$300$ volume of the IllustrisTNG project \citep{2018Springel+, 2018Marinacci+, 2018Nelson+, 2018Pillepich+, 2018Naiman+} and the recent TNG-Cluster simulations \citep{2024Nelson+}, with slight differences arising from the choice of cosmology. 

We apply a physics model based on that used in the TNG project, which is described in detail in \citet{2017Weinberger+} and \citet{2018PillepichMethod+}. Physical processes in this model include primordial and metal line cooling along with heating from a spatially uniform UV background \citep{2013Vogelsberger+}, an effective model for the ISM and star formation \citep{2003SpringelHernquist}, chemical enrichment of the ISM due to core collapse and thermonuclear supernovae and AGB stars, an effective model for galactic winds \citep{2018PillepichMethod+} and the formation, growth of and feedback from supermassive black holes \citep{2017Weinberger+}. We additionally initialise a spatially uniform seed magnetic field of comoving strength of $10^{-14} \, {\rm G}$ at the start of the simulations ($z = 127$). 

We perform three zoom simulations of this halo, all of which use this physics model as a baseline. The first simulation (which we will refer to as the `standard' simulation) has no additional models for physical processes. In the other two simulations we switch on our model for anisotropic thermal conduction. In both of these simulations we use the `semi-implicit, linear solver'. In one of these simulations, we set the conductivity to the Spitzer value (equation~\ref{eq: spitzer}) and in the other we set it to Spitzer with whistler suppression (equation~\ref{eq: kappa interp}). We refer to these simulations as `+spitzer' and `+whistler', respectively.

\begin{figure*}
    \centering
    \includegraphics[width=\textwidth]{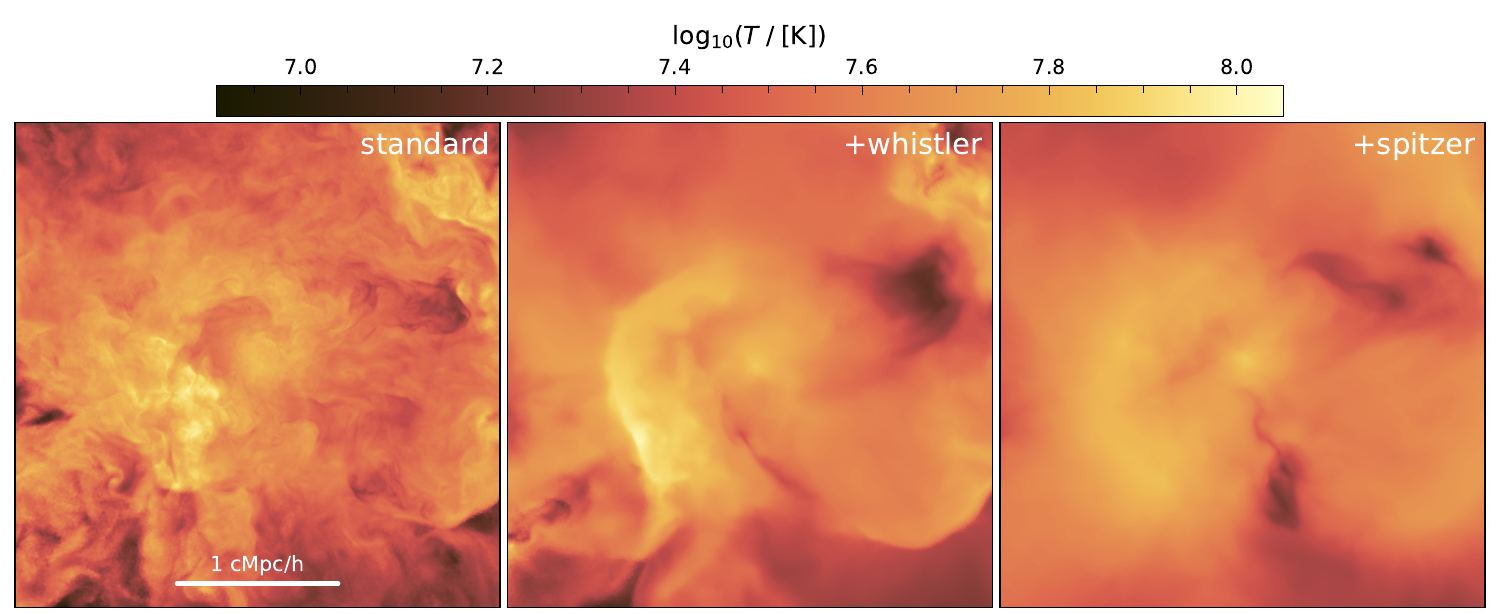}
    \caption[Cluster temp variations projections]
    {Mass-weighted temperature projections of size $(3\,h^{-1} {\rm Mpc})^2$ and depth $200 \,h^{-1} {\rm kpc}$ at $z=0$ for, from left to right, the simulation without conduction (`standard'), that with whistler suppressed anisotropic conduction (`+whistler'), and that with full Spitzer conductivity (`+spitzer'). The `standard' simulation shows significant small-scale structure in the temperature field. The temperature field in the simulations with conduction, however, is considerably smoother, with the largest effect seen in the `+spitzer' case.}
    \label{fig: cluster var temp}
\end{figure*}

\begin{figure}
    \centering
    \includegraphics[width=0.49\textwidth]{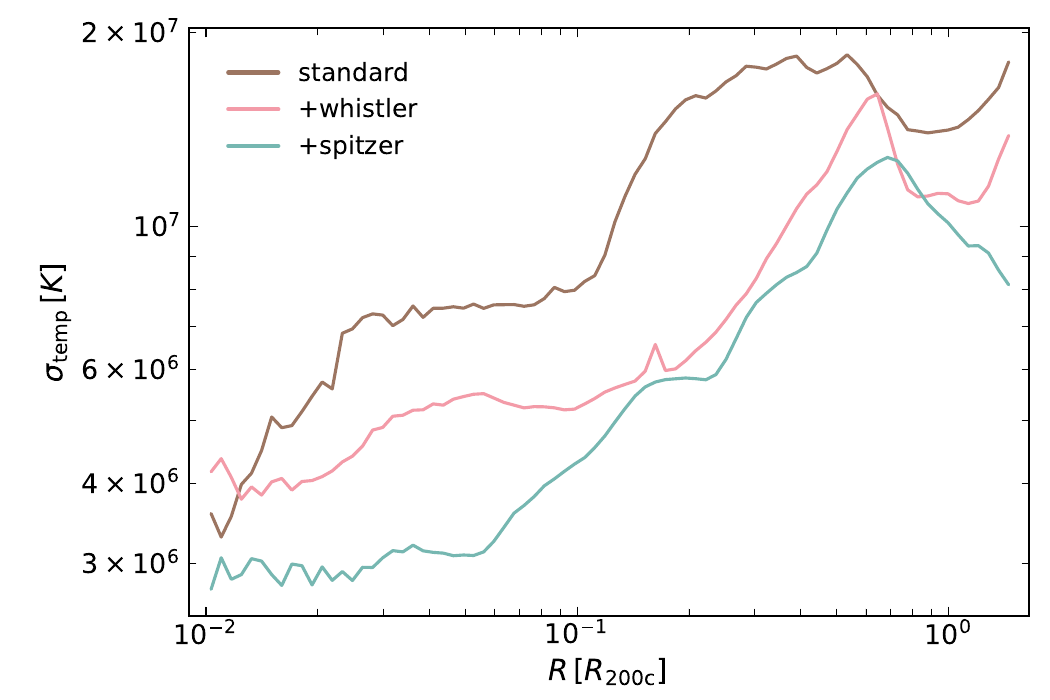}
    \caption[Cluster temp variations profile]
    {Radial profiles at $z=0$ of the standard deviation of the temperature distribution in radial bins, weighted by volume, for the simulation without conduction (`standard'), that with whistler suppressed anisotropic conduction (`+whistler'), and that with full Spitzer conductivity (`+spitzer'). The `standard' simulation shows the highest levels of variability in the temperature field, whereas the levels of variability in both simulations with anisotropic conduction are significantly lower.}
    \label{fig: cluster var profile}
\end{figure}

For the simulations with thermal conduction, we do not solve the conduction problem across interfaces where one or more of the cells on either side are star forming. The thermal energy of these cells is set by the effective equation of state \citep{2003SpringelHernquist} and represents an effective pressure rather than the temperature of the gas, which should just be considered as a property of the effective subgrid model. In doing this we effectively assume that thermal conduction is unimportant for the transport of energy between the star-forming ISM and the gas surrounding it.

\subsection{Results}

Figure~\ref{fig: cluster viz} shows thin projections of various quantities at $z = 0$ for the `+whistler' simulation. The magnetic field in the cluster is clearly highly turbulent and has strength that reaches $\sim 10 \, \mu {\rm G}$ in the central regions. In the outskirts of the cluster, however, the field is significantly weaker. For an in-depth discussion of the amplification of magnetic fields in the PICO clusters, see Tevlin et al. to be submitted.

Throughout the majority of the cluster, the magnetic field is, however, largely dynamically unimportant, as can be seen in the projection of the plasma beta. The plasma beta typically lies within the range $\sim10^2-10^3$ but can be even higher in the cluster outskirts, particularly in regions where the magnetic field is low. Intermittent magnetic flux tubes can reach plasma beta values of 10.

In the lower-middle panel, we show the thermal diffusivity $\kappa = \chi / (\rho c_{\rm v})$, as measured in the simulation (i.e.~including the effects of whistler suppression) and in the lower-right panel we show the factor by which the diffusivity is suppressed below the Spitzer value, $(1 + (1/3) \,\beta \, \lambda_{\rm mfp, e} \, / \, l_{\rm T, \parallel})^{-1}$, see equation~(\ref{eq: kappa interp}). The turbulent nature of the magnetic field (in combination with a much smoother temperature gradient) leads to significant variation in the temperature gradient length scale along the direction of the magnetic field, $l_{\rm T, \parallel} = |\boldsymbol{b} \boldsymbol{\cdot} \boldsymbol{\nabla} \ln T|^{-1}$. This, in combination with the variations in $\beta$, can lead to a significant suppression of the conductivity, particularly in the outskirts of the cluster and at the location of shocks (see the temperature projection in the upper-left panel and density in the upper-middle panel). 

In Fig.~\ref{fig: cluster profiles} we show radially averaged profiles of various thermodynamical quantities at $z=0$ for all three simulations. In general, we find good agreement across all three runs. There are, however, some slight differences, particularly in  the density, metallicity, entropy and magnetic field strength profiles: the runs with the conduction exhibit higher central densities and magnetic field strengths, lower central entropies and flatter central metallicity gradients. Additionally, the cluster simulated with full Spitzer conductivity shows larger differences than that simulated with whistler suppressed conduction when compared to the simulation without conduction.

Whilst the differences in the radially averaged profiles are rather subtle, we do find larger differences between these systems in other quantities. In Fig.~\ref{fig: cluster var temp} we show temperature projections of the three clusters at $z=0$. The `standard' simulation shows significant small-scale structure in the temperature field. In the simulations with conduction, however, this small-scale structure is largely gone and the temperature field is substantially smoother; an effect that is most pronounced in the `+spitzer' run, where the conductivities are generally higher than in the `+whistler' run (see lower right panel of Fig.~\ref{fig: cluster profiles}). These observations are consistent with the expected behaviour of thermal conduction: namely that it acts to flatten temperature gradients.

This can be understood more quantitatively by examining radial profiles of the standard deviation of the temperature within each radial bin, weighted by cell volume, which are shown in Fig.~\ref{fig: cluster var profile}. At all radii (except the very centre) the simulation without conduction has the largest temperature variance, while both simulations with anisotropic conduction show significantly lower levels of variability. The `+spitzer' simulation, where the conductivities are generally highest, has the lowest level of variability. 

In this section we have shown that the differences between simulations with and without anisotropic conduction can be significant. Additionally, we have shown that the effect of whistler suppression can be considerable. Previous cosmological cluster simulations have also found that conduction can have a substantial effect on the cluster properties \citep[see e.g.][]{2004Dolag+, Jubelgas2004, 2011Ruszkowski+, 2017Kannan+, 2019Barnes+, 2023Pellissier+}. In order to carefully asses the extent to which thermal conduction is affecting the cluster properties and in what ways these effects manifest, it is first important to better understand the variability inherent in the standard physics model and which cluster properties it robustly predicts. It is, additionally, important to have more than one data point to draw conclusions from, which requires simulating a larger sample of clusters. This is beyond the scope of this work, where the aim is to introduce our thermal conduction scheme, but will be studied in detail in future work.

It should also be mentioned that the extent to which the thermal conductivity closure used in the `+whistler' simulation (equation~\ref{eq: kappa interp}) accurately captures the effect of whistler waves on electron transport in the ICM is not well understood. The present results, however, demonstrate that it may have a significant impact on the effective conductivity, particularly in the outskirts of the cluster. Future work is needed to better understand the impact of kinetic microinstabilities on electron transport and how these effects manifest on large scales, in a cosmological context.

\subsection{Technical considerations}

We end this section with a few technical remarks relating to the performance of the anisotropic thermal conduction solver in these cluster zooms. In the two simulations with thermal conduction, about $20\%$ of the run-time is spent in the conduction solver. About a third to half of this time is spent in the matrix solver, and the majority of the remaining time is spent in the computation of the coefficients for the gradient estimates and the preparation of the relevant data structures. The cost associated with the explicit flux calculation itself is negligible in comparison. 

Fewer than $10$ iterations are typically required to solve the system of equations in the implicit part of the calculation. At later times, and particularly at time steps where a significant fraction of the gas cells are active, the addition of the multigrid preconditioner is required as otherwise the matrix solver is unable to converge within $200$ iterations. 

As is typical for cluster zooms, our simulations have timestep hierarchies that can be up to $\sim 10$ levels deep. These simulations would be prohibitively expensive if all cells had to be integrated on the smallest global timestep. The significant difference between the smallest and largest timesteps (a factor of $\sim1024$) demonstrates that it is not appropriate to apply the scheme only on global synchronisation points. All this underlines the necessity of our local timestepping scheme for such calculations.

The data structures that store necessary information related to the faces and corners of the Voronoi mesh have the largest memory footprint of all those associated with the conduction solver. In addition, the {\small HYPRE} library can have a significant memory footprint in larger simulations which is allocated independently to {\small AREPO}'s memory manager. All these structures are, however, not required anywhere else in the code, and are freed upon exit of the thermal conduction routine.

\section{Summary and outlook}
\label{sec: summary}
In this paper, we presented a new scheme for modelling anisotropic thermal conduction on a moving mesh. The solver, which we have implemented into the moving-mesh code {\small AREPO} \citep{2010Springel, 2016PakmorNum, 2020Weinberger+}, is fully conservative, ensures the entropy condition is not violated, and allows for semi-implicit time integration and individual timestepping. 

The anisotropic thermal conduction solver introduced in this work represents an improvement over the approach described in \citet{2016Kannan+}, which is also implemented in {\small AREPO}, primarily due to the fact that it supports local timestepping; a feature which is crucial for efficiency and accuracy in cosmological simulations. If conduction is done on global timesteps only, it either becomes too expensive to run, or one loses the coupling to the gas on faster timescales. Accurate treatment of anisotropic thermal conduction is, additionally, more complex than cosmic ray diffusion \citep{2016Pakmor+}, due to the fact that the conductivity is not spatially and temporally constant (as is often assumed in one-moment treatments of cosmic ray diffusion). 

We tested our implementation on a variety of numerical problems and demonstrated that our solver is able to reproduce analytic predictions and numerical solutions. The convergence rate of the anisotropic solver is comparable to those of other numerical diffusion solvers, and we showed that the use of local timesteps does not lead to a significant reduction in the accuracy of the solver. We also verified the accuracy of the coupling between the conduction and hydrodynamics and demonstrated that the solver is able to reproduce the speed of the conduction front predicted by analytic arguments. These tests highlighted just a few ways anisotropic thermal conduction can fundamentally change the behaviour of a system.

We also demonstrated that the solver can be applied to highly non-linear problems with deep timestep hierarchies by performing high-resolution cosmological zoom-in simulations of a galaxy cluster with anisotropic conduction, where we showed that the solver operates efficiently and robustly. In these simulations we found that anisotropic thermal conduction, as well as the presence (or absence) of whistler suppression can have a significant impact on the temperature distribution of ICM and acts to smooth out small-scale structure. We will explore these results in more detail in future work.

The ability to accurately and efficiently capture the effects of anisotropic thermal conduction in a fully cosmological environment is crucial, and the methods presented here will allow us to assess the effects of conduction in a wide range of astrophysical processes in diverse contexts. 

In the future, we plan to use high resolution cosmological simulations to better understand the role anisotropic thermal conduction plays in a range of astrophysical contexts, including the ICM of clusters and the CGM of galaxies. This will require a large suite of high resolution cosmological zoom simulations. These studies will be made possible by the fact that our conduction solver is able to operate accurately and efficiently on problems that exhibit significant variations in the relevant spatial and temporal scales.

\section*{Acknowledgements}
We would like to thank Lorenzo Maria Perrone for very helpful discussions and suggestions. CP acknowledges support by the European Research Council under ERC-AdG grant PICOGAL-101019746. RB is supported by the UZH Postdoc Grant, grant no. FK-23116 and the SNSF through the Ambizione Grant PZ00P2\_223532. FvdV is supported by a Royal Society University Research Fellowship (URF\textbackslash R1\textbackslash 191703).

\section*{Data Availability}
The data underlying this article will be shared upon request to the corresponding author.



\bibliographystyle{mnras}
\bibliography{references} 

\begin{thebibliography}{}
\makeatletter
\relax
\def\mn@urlcharsother{\let\do\@makeother \do\$\do\&\do\#\do\^\do\_\do\%\do\~}
\def\mn@doi{\begingroup\mn@urlcharsother \@ifnextchar [ {\mn@doi@} {\mn@doi@[]}}
\def\mn@doi@[#1]#2{\def\@tempa{#1}\ifx\@tempa\@empty \href {http://dx.doi.org/#2} {doi:#2}\else \href {http://dx.doi.org/#2} {#1}\fi \endgroup}
\def\mn@eprint#1#2{\mn@eprint@#1:#2::\@nil}
\def\mn@eprint@arXiv#1{\href {http://arxiv.org/abs/#1} {{\tt arXiv:#1}}}
\def\mn@eprint@dblp#1{\href {http://dblp.uni-trier.de/rec/bibtex/#1.xml} {dblp:#1}}
\def\mn@eprint@#1:#2:#3:#4\@nil{\def\@tempa {#1}\def\@tempb {#2}\def\@tempc {#3}\ifx \@tempc \@empty \let \@tempc \@tempb \let \@tempb \@tempa \fi \ifx \@tempb \@empty \def\@tempb {arXiv}\fi \@ifundefined {mn@eprint@\@tempb}{\@tempb:\@tempc}{\expandafter \expandafter \csname mn@eprint@\@tempb\endcsname \expandafter{\@tempc}}}

\bibitem[\protect\citeauthoryear{{Armillotta}, {Fraternali}, {Werk}, {Prochaska}  \& {Marinacci}}{{Armillotta} et~al.}{2017}]{2017Armillotta+}
{Armillotta} L.,  {Fraternali} F.,  {Werk} J.~K.,  {Prochaska} J.~X.,   {Marinacci} F.,  2017, \mn@doi [\mnras] {10.1093/mnras/stx1239}, \href {https://ui.adsabs.harvard.edu/abs/2017MNRAS.470..114A} {470, 114}

\bibitem[\protect\citeauthoryear{{Asai}, {Fukuda}  \& {Matsumoto}}{{Asai} et~al.}{2007}]{Asai2007}
{Asai} N.,  {Fukuda} N.,   {Matsumoto} R.,  2007, \mn@doi [\apj] {10.1086/518235}, \href {https://ui.adsabs.harvard.edu/abs/2007ApJ...663..816A} {663, 816}

\bibitem[\protect\citeauthoryear{{Balbus}}{{Balbus}}{2000}]{2000Balbus}
{Balbus} S.~A.,  2000, \mn@doi [\apj] {10.1086/308732}, \href {https://ui.adsabs.harvard.edu/abs/2000ApJ...534..420B} {534, 420}

\bibitem[\protect\citeauthoryear{{Balbus}}{{Balbus}}{2001}]{2001Balbus}
{Balbus} S.~A.,  2001, \mn@doi [\apj] {10.1086/323875}, \href {https://ui.adsabs.harvard.edu/abs/2001ApJ...562..909B} {562, 909}

\bibitem[\protect\citeauthoryear{{Balbus} \& {Reynolds}}{{Balbus} \& {Reynolds}}{2010}]{Balbus2010}
{Balbus} S.~A.,  {Reynolds} C.~S.,  2010, \mn@doi [\apjl] {10.1088/2041-8205/720/1/L97}, \href {https://ui.adsabs.harvard.edu/abs/2010ApJ...720L..97B} {720, L97}

\bibitem[\protect\citeauthoryear{{Balsara}, {Tilley}  \& {Howk}}{{Balsara} et~al.}{2008a}]{2008Balsara+}
{Balsara} D.~S.,  {Tilley} D.~A.,   {Howk} J.~C.,  2008a, \mn@doi [\mnras] {10.1111/j.1365-2966.2008.13085.x}, \href {https://ui.adsabs.harvard.edu/abs/2008MNRAS.386..627B} {386, 627}

\bibitem[\protect\citeauthoryear{{Balsara}, {Bendinelli}, {Tilley}, {Massari}  \& {Howk}}{{Balsara} et~al.}{2008b}]{2008Balsarab+}
{Balsara} D.~S.,  {Bendinelli} A.~J.,  {Tilley} D.~A.,  {Massari} A.~R.,   {Howk} J.~C.,  2008b, \mn@doi [\mnras] {10.1111/j.1365-2966.2008.13121.x}, \href {https://ui.adsabs.harvard.edu/abs/2008MNRAS.386..642B} {386, 642}

\bibitem[\protect\citeauthoryear{{Barenblatt}}{{Barenblatt}}{1996}]{1996Barrenblatt}
{Barenblatt} G.~I.,  1996, {Scaling, Self-similarity, and Intermediate Asymptotics}

\bibitem[\protect\citeauthoryear{{Barnes} et~al.,}{{Barnes} et~al.}{2019}]{2019Barnes+}
{Barnes} D.~J.,  et~al., 2019, \mn@doi [\mnras] {10.1093/mnras/stz1814}, \href {https://ui.adsabs.harvard.edu/abs/2019MNRAS.488.3003B} {488, 3003}

\bibitem[\protect\citeauthoryear{{Beckmann}, {Dubois}, {Pellissier}, {Polles}  \& {Olivares}}{{Beckmann} et~al.}{2022}]{2022Beckmann+}
{Beckmann} R.~S.,  {Dubois} Y.,  {Pellissier} A.,  {Polles} F.~L.,   {Olivares} V.,  2022, \mn@doi [\aap] {10.1051/0004-6361/202243873}, \href {https://ui.adsabs.harvard.edu/abs/2022A&A...666A..71B} {666, A71}

\bibitem[\protect\citeauthoryear{{Berlok}, {Pakmor}  \& {Pfrommer}}{{Berlok} et~al.}{2020}]{2020Berlok+}
{Berlok} T.,  {Pakmor} R.,   {Pfrommer} C.,  2020, \mn@doi [\mnras] {10.1093/mnras/stz3115}, \href {https://ui.adsabs.harvard.edu/abs/2020MNRAS.491.2919B} {491, 2919}

\bibitem[\protect\citeauthoryear{{Berlok}, {Quataert}, {Pessah}  \& {Pfrommer}}{{Berlok} et~al.}{2021}]{2021Berlok+}
{Berlok} T.,  {Quataert} E.,  {Pessah} M.~E.,   {Pfrommer} C.,  2021, \mn@doi [\mnras] {10.1093/mnras/stab832}, \href {https://ui.adsabs.harvard.edu/abs/2021MNRAS.504.3435B} {504, 3435}

\bibitem[\protect\citeauthoryear{{Bingert} \& {Peter}}{{Bingert} \& {Peter}}{2011}]{2011BingertPeter}
{Bingert} S.,  {Peter} H.,  2011, \mn@doi [\aap] {10.1051/0004-6361/201016019}, \href {https://ui.adsabs.harvard.edu/abs/2011A&A...530A.112B} {530, A112}

\bibitem[\protect\citeauthoryear{{Bourdin}, {Bingert}  \& {Peter}}{{Bourdin} et~al.}{2013}]{2013Bourdin+}
{Bourdin} P.~A.,  {Bingert} S.,   {Peter} H.,  2013, \mn@doi [\aap] {10.1051/0004-6361/201321185}, \href {https://ui.adsabs.harvard.edu/abs/2013A&A...555A.123B} {555, A123}

\bibitem[\protect\citeauthoryear{{Braginskii}}{{Braginskii}}{1965}]{1965Braginskii}
{Braginskii} S.~I.,  1965, Reviews of Plasma Physics, \href {https://ui.adsabs.harvard.edu/abs/1965RvPP....1..205B} {1, 205}

\bibitem[\protect\citeauthoryear{{Br{\"u}ggen} \& {Scannapieco}}{{Br{\"u}ggen} \& {Scannapieco}}{2016}]{2016BrueggenScannapieco}
{Br{\"u}ggen} M.,  {Scannapieco} E.,  2016, \mn@doi [\apj] {10.3847/0004-637X/822/1/31}, \href {https://ui.adsabs.harvard.edu/abs/2016ApJ...822...31B} {822, 31}

\bibitem[\protect\citeauthoryear{{Br{\"u}ggen}, {Scannapieco}  \& {Grete}}{{Br{\"u}ggen} et~al.}{2023}]{2023BrueggenScannapieco}
{Br{\"u}ggen} M.,  {Scannapieco} E.,   {Grete} P.,  2023, \mn@doi [\apj] {10.3847/1538-4357/acd63e}, \href {https://ui.adsabs.harvard.edu/abs/2023ApJ...951..113B} {951, 113}

\bibitem[\protect\citeauthoryear{{Chevalier}}{{Chevalier}}{1975}]{1975Chevalier}
{Chevalier} R.~A.,  1975, \mn@doi [\apj] {10.1086/153840}, \href {https://ui.adsabs.harvard.edu/abs/1975ApJ...200..698C} {200, 698}

\bibitem[\protect\citeauthoryear{{Choi} \& {Stone}}{{Choi} \& {Stone}}{2012}]{2012ChoiStone}
{Choi} E.,  {Stone} J.~M.,  2012, \mn@doi [\apj] {10.1088/0004-637X/747/2/86}, \href {https://ui.adsabs.harvard.edu/abs/2012ApJ...747...86C} {747, 86}

\bibitem[\protect\citeauthoryear{{Cowie} \& {McKee}}{{Cowie} \& {McKee}}{1977}]{1977CowieMcKee}
{Cowie} L.~L.,  {McKee} C.~F.,  1977, \mn@doi [\apj] {10.1086/154911}, \href {https://ui.adsabs.harvard.edu/abs/1977ApJ...211..135C} {211, 135}

\bibitem[\protect\citeauthoryear{{Dolag}, {Jubelgas}, {Springel}, {Borgani}  \& {Rasia}}{{Dolag} et~al.}{2004}]{2004Dolag+}
{Dolag} K.,  {Jubelgas} M.,  {Springel} V.,  {Borgani} S.,   {Rasia} E.,  2004, \mn@doi [\apjl] {10.1086/420966}, \href {https://ui.adsabs.harvard.edu/abs/2004ApJ...606L..97D} {606, L97}

\bibitem[\protect\citeauthoryear{{Dong} \& {Stone}}{{Dong} \& {Stone}}{2009}]{2009DongStone}
{Dong} R.,  {Stone} J.~M.,  2009, \mn@doi [\apj] {10.1088/0004-637X/704/2/1309}, \href {https://ui.adsabs.harvard.edu/abs/2009ApJ...704.1309D} {704, 1309}

\bibitem[\protect\citeauthoryear{{Drake} et~al.,}{{Drake} et~al.}{2021}]{Drake2021}
{Drake} J.~F.,  et~al., 2021, \mn@doi [\apj] {10.3847/1538-4357/ac1ff1}, \href {https://ui.adsabs.harvard.edu/abs/2021ApJ...923..245D} {923, 245}

\bibitem[\protect\citeauthoryear{{Dubois} \& {Commer{\c{c}}on}}{{Dubois} \& {Commer{\c{c}}on}}{2016}]{2016DuboisCommercon}
{Dubois} Y.,  {Commer{\c{c}}on} B.,  2016, \mn@doi [\aap] {10.1051/0004-6361/201527126}, \href {https://ui.adsabs.harvard.edu/abs/2016A&A...585A.138D} {585, A138}

\bibitem[\protect\citeauthoryear{{Dursi} \& {Pfrommer}}{{Dursi} \& {Pfrommer}}{2008}]{2008DursiPfrommer}
{Dursi} L.~J.,  {Pfrommer} C.,  2008, \mn@doi [\apj] {10.1086/529371}, \href {https://ui.adsabs.harvard.edu/abs/2008ApJ...677..993D} {677, 993}

\bibitem[\protect\citeauthoryear{{Fabian}, {Sanders}, {Crawford}, {Conselice}, {Gallagher}  \& {Wyse}}{{Fabian} et~al.}{2003}]{2003Fabian+}
{Fabian} A.~C.,  {Sanders} J.~S.,  {Crawford} C.~S.,  {Conselice} C.~J.,  {Gallagher} J.~S.,   {Wyse} R.~F.~G.,  2003, \mn@doi [\mnras] {10.1046/j.1365-8711.2003.06856.x}, \href {https://ui.adsabs.harvard.edu/abs/2003MNRAS.344L..48F} {344, L48}

\bibitem[\protect\citeauthoryear{Falgout \& Yang}{Falgout \& Yang}{2002}]{2002FalgoutYang}
Falgout R.~D.,  Yang U.~M.,  2002, in Sloot P. M.~A.,  Hoekstra A.~G.,  Tan C. J.~K.,   Dongarra J.~J.,  eds, Computational Science --- ICCS 2002. Springer Berlin Heidelberg, Berlin, Heidelberg, pp 632--641

\bibitem[\protect\citeauthoryear{Henson \& Yang}{Henson \& Yang}{2002}]{2002HensonYang}
Henson V.,  Yang U.,  2002, \mn@doi [Applied Numerical Mathematics] {10.1016/S0168-9274(01)00115-5}, 41, 155

\bibitem[\protect\citeauthoryear{{Jacob} \& {Pfrommer}}{{Jacob} \& {Pfrommer}}{2017a}]{2017JacobPfrommerA}
{Jacob} S.,  {Pfrommer} C.,  2017a, \mn@doi [\mnras] {10.1093/mnras/stx131}, \href {https://ui.adsabs.harvard.edu/abs/2017MNRAS.467.1449J} {467, 1449}

\bibitem[\protect\citeauthoryear{{Jacob} \& {Pfrommer}}{{Jacob} \& {Pfrommer}}{2017b}]{2017JacobPfrommerB}
{Jacob} S.,  {Pfrommer} C.,  2017b, \mn@doi [\mnras] {10.1093/mnras/stx132}, \href {https://ui.adsabs.harvard.edu/abs/2017MNRAS.467.1478J} {467, 1478}

\bibitem[\protect\citeauthoryear{{Jennings} \& {Li}}{{Jennings} \& {Li}}{2021}]{2021JenningsLi}
{Jennings} R.~M.,  {Li} Y.,  2021, \mn@doi [\mnras] {10.1093/mnras/stab1607}, \href {https://ui.adsabs.harvard.edu/abs/2021MNRAS.505.5238J} {505, 5238}

\bibitem[\protect\citeauthoryear{{Jubelgas}, {Springel}  \& {Dolag}}{{Jubelgas} et~al.}{2004}]{Jubelgas2004}
{Jubelgas} M.,  {Springel} V.,   {Dolag} K.,  2004, \mn@doi [\mnras] {10.1111/j.1365-2966.2004.07801.x}, \href {https://ui.adsabs.harvard.edu/abs/2004MNRAS.351..423J} {351, 423}

\bibitem[\protect\citeauthoryear{{Kannan}, {Springel}, {Pakmor}, {Marinacci}  \& {Vogelsberger}}{{Kannan} et~al.}{2016}]{2016Kannan+}
{Kannan} R.,  {Springel} V.,  {Pakmor} R.,  {Marinacci} F.,   {Vogelsberger} M.,  2016, \mn@doi [\mnras] {10.1093/mnras/stw294}, \href {https://ui.adsabs.harvard.edu/abs/2016MNRAS.458..410K} {458, 410}

\bibitem[\protect\citeauthoryear{{Kannan}, {Vogelsberger}, {Pfrommer}, {Weinberger}, {Springel}, {Hernquist}, {Puchwein}  \& {Pakmor}}{{Kannan} et~al.}{2017}]{2017Kannan+}
{Kannan} R.,  {Vogelsberger} M.,  {Pfrommer} C.,  {Weinberger} R.,  {Springel} V.,  {Hernquist} L.,  {Puchwein} E.,   {Pakmor} R.,  2017, \mn@doi [\apjl] {10.3847/2041-8213/aa624b}, \href {https://ui.adsabs.harvard.edu/abs/2017ApJ...837L..18K} {837, L18}

\bibitem[\protect\citeauthoryear{{Komarov}, {Churazov}, {Kunz}  \& {Schekochihin}}{{Komarov} et~al.}{2016}]{2016Komarov+}
{Komarov} S.~V.,  {Churazov} E.~M.,  {Kunz} M.~W.,   {Schekochihin} A.~A.,  2016, \mn@doi [\mnras] {10.1093/mnras/stw963}, \href {https://ui.adsabs.harvard.edu/abs/2016MNRAS.460..467K} {460, 467}

\bibitem[\protect\citeauthoryear{{Komarov}, {Schekochihin}, {Churazov}  \& {Spitkovsky}}{{Komarov} et~al.}{2018}]{2018Komarov+}
{Komarov} S.,  {Schekochihin} A.~A.,  {Churazov} E.,   {Spitkovsky} A.,  2018, \mn@doi [Journal of Plasma Physics] {10.1017/S0022377818000399}, \href {https://ui.adsabs.harvard.edu/abs/2018JPlPh..84c9005K} {84, 905840305}

\bibitem[\protect\citeauthoryear{{Kunz}}{{Kunz}}{2011}]{2011Kunz}
{Kunz} M.~W.,  2011, \mn@doi [\mnras] {10.1111/j.1365-2966.2011.19303.x}, \href {https://ui.adsabs.harvard.edu/abs/2011MNRAS.417..602K} {417, 602}

\bibitem[\protect\citeauthoryear{{Kunz}, {Schekochihin}  \& {Stone}}{{Kunz} et~al.}{2014}]{2014Kunz+}
{Kunz} M.~W.,  {Schekochihin} A.~A.,   {Stone} J.~M.,  2014, \mn@doi [\prl] {10.1103/PhysRevLett.112.205003}, \href {https://ui.adsabs.harvard.edu/abs/2014PhRvL.112t5003K} {112, 205003}

\bibitem[\protect\citeauthoryear{{Lyutikov}}{{Lyutikov}}{2006}]{Lyutikov2006}
{Lyutikov} M.,  2006, \mn@doi [\mnras] {10.1111/j.1365-2966.2006.10835.x}, \href {https://ui.adsabs.harvard.edu/abs/2006MNRAS.373...73L} {373, 73}

\bibitem[\protect\citeauthoryear{{Marinacci} et~al.,}{{Marinacci} et~al.}{2018}]{2018Marinacci+}
{Marinacci} F.,  et~al., 2018, \mn@doi [\mnras] {10.1093/mnras/sty2206}, \href {https://ui.adsabs.harvard.edu/abs/2018MNRAS.480.5113M} {480, 5113}

\bibitem[\protect\citeauthoryear{{Markevitch} \& {Vikhlinin}}{{Markevitch} \& {Vikhlinin}}{2007}]{2007MarkevitchVikhlinin}
{Markevitch} M.,  {Vikhlinin} A.,  2007, \mn@doi [\physrep] {10.1016/j.physrep.2007.01.001}, \href {https://ui.adsabs.harvard.edu/abs/2007PhR...443....1M} {443, 1}

\bibitem[\protect\citeauthoryear{{McCourt}, {Parrish}, {Sharma}  \& {Quataert}}{{McCourt} et~al.}{2011}]{2011McCourt+}
{McCourt} M.,  {Parrish} I.~J.,  {Sharma} P.,   {Quataert} E.,  2011, \mn@doi [\mnras] {10.1111/j.1365-2966.2011.18216.x}, \href {https://ui.adsabs.harvard.edu/abs/2011MNRAS.413.1295M} {413, 1295}

\bibitem[\protect\citeauthoryear{{Naiman} et~al.,}{{Naiman} et~al.}{2018}]{2018Naiman+}
{Naiman} J.~P.,  et~al., 2018, \mn@doi [\mnras] {10.1093/mnras/sty618}, \href {https://ui.adsabs.harvard.edu/abs/2018MNRAS.477.1206N} {477, 1206}

\bibitem[\protect\citeauthoryear{{Navarro}, {Khomenko}, {Modestov}  \& {Vitas}}{{Navarro} et~al.}{2022}]{2022Navarro+}
{Navarro} A.,  {Khomenko} E.,  {Modestov} M.,   {Vitas} N.,  2022, \mn@doi [\aap] {10.1051/0004-6361/202243439}, \href {https://ui.adsabs.harvard.edu/abs/2022A&A...663A..96N} {663, A96}

\bibitem[\protect\citeauthoryear{{Nelson} et~al.,}{{Nelson} et~al.}{2018}]{2018Nelson+}
{Nelson} D.,  et~al., 2018, \mn@doi [\mnras] {10.1093/mnras/stx3040}, \href {https://ui.adsabs.harvard.edu/abs/2018MNRAS.475..624N} {475, 624}

\bibitem[\protect\citeauthoryear{{Nelson}, {Pillepich}, {Ayromlou}, {Lee}, {Lehle}, {Rohr}  \& {Truong}}{{Nelson} et~al.}{2024}]{2024Nelson+}
{Nelson} D.,  {Pillepich} A.,  {Ayromlou} M.,  {Lee} W.,  {Lehle} K.,  {Rohr} E.,   {Truong} N.,  2024, \mn@doi [\aap] {10.1051/0004-6361/202348608}, \href {https://ui.adsabs.harvard.edu/abs/2024A&A...686A.157N} {686, A157}

\bibitem[\protect\citeauthoryear{{Pakmor} \& {Springel}}{{Pakmor} \& {Springel}}{2013}]{2013PakmorSpringel}
{Pakmor} R.,  {Springel} V.,  2013, \mn@doi [\mnras] {10.1093/mnras/stt428}, \href {https://ui.adsabs.harvard.edu/abs/2013MNRAS.432..176P} {432, 176}

\bibitem[\protect\citeauthoryear{{Pakmor}, {Bauer}  \& {Springel}}{{Pakmor} et~al.}{2011}]{2011Pakmor+}
{Pakmor} R.,  {Bauer} A.,   {Springel} V.,  2011, \mn@doi [\mnras] {10.1111/j.1365-2966.2011.19591.x}, \href {https://ui.adsabs.harvard.edu/abs/2011MNRAS.418.1392P} {418, 1392}

\bibitem[\protect\citeauthoryear{{Pakmor}, {Springel}, {Bauer}, {Mocz}, {Munoz}, {Ohlmann}, {Schaal}  \& {Zhu}}{{Pakmor} et~al.}{2016a}]{2016PakmorNum}
{Pakmor} R.,  {Springel} V.,  {Bauer} A.,  {Mocz} P.,  {Munoz} D.~J.,  {Ohlmann} S.~T.,  {Schaal} K.,   {Zhu} C.,  2016a, \mn@doi [\mnras] {10.1093/mnras/stv2380}, \href {https://ui.adsabs.harvard.edu/abs/2016MNRAS.455.1134P} {455, 1134}

\bibitem[\protect\citeauthoryear{{Pakmor}, {Pfrommer}, {Simpson}, {Kannan}  \& {Springel}}{{Pakmor} et~al.}{2016b}]{2016Pakmor+}
{Pakmor} R.,  {Pfrommer} C.,  {Simpson} C.~M.,  {Kannan} R.,   {Springel} V.,  2016b, \mn@doi [\mnras] {10.1093/mnras/stw1761}, \href {https://ui.adsabs.harvard.edu/abs/2016MNRAS.462.2603P} {462, 2603}

\bibitem[\protect\citeauthoryear{{Pakmor} et~al.,}{{Pakmor} et~al.}{2023}]{2023Pakmor+}
{Pakmor} R.,  et~al., 2023, \mn@doi [\mnras] {10.1093/mnras/stac3620}, \href {https://ui.adsabs.harvard.edu/abs/2023MNRAS.524.2539P} {524, 2539}

\bibitem[\protect\citeauthoryear{{Parrish} \& {Quataert}}{{Parrish} \& {Quataert}}{2008}]{2008ParrishQuataert}
{Parrish} I.~J.,  {Quataert} E.,  2008, \mn@doi [\apjl] {10.1086/587937}, \href {https://ui.adsabs.harvard.edu/abs/2008ApJ...677L...9P} {677, L9}

\bibitem[\protect\citeauthoryear{{Parrish} \& {Stone}}{{Parrish} \& {Stone}}{2005}]{2005ParrishStone}
{Parrish} I.~J.,  {Stone} J.~M.,  2005, \mn@doi [\apj] {10.1086/444589}, \href {https://ui.adsabs.harvard.edu/abs/2005ApJ...633..334P} {633, 334}

\bibitem[\protect\citeauthoryear{{Parrish} \& {Stone}}{{Parrish} \& {Stone}}{2007}]{2007ParrishStone}
{Parrish} I.~J.,  {Stone} J.~M.,  2007, \mn@doi [\apj] {10.1086/518881}, \href {https://ui.adsabs.harvard.edu/abs/2007ApJ...664..135P} {664, 135}

\bibitem[\protect\citeauthoryear{{Parrish}, {Quataert}  \& {Sharma}}{{Parrish} et~al.}{2010}]{Parrish2010}
{Parrish} I.~J.,  {Quataert} E.,   {Sharma} P.,  2010, \mn@doi [\apjl] {10.1088/2041-8205/712/2/L194}, \href {https://ui.adsabs.harvard.edu/abs/2010ApJ...712L.194P} {712, L194}

\bibitem[\protect\citeauthoryear{{Pellissier}, {Hahn}  \& {Ferrari}}{{Pellissier} et~al.}{2023}]{2023Pellissier+}
{Pellissier} A.,  {Hahn} O.,   {Ferrari} C.,  2023, \mn@doi [\mnras] {10.1093/mnras/stad888}, \href {https://ui.adsabs.harvard.edu/abs/2023MNRAS.522..721P} {522, 721}

\bibitem[\protect\citeauthoryear{{Perrone} \& {Latter}}{{Perrone} \& {Latter}}{2022a}]{Perrone2022a}
{Perrone} L.~M.,  {Latter} H.,  2022a, \mn@doi [\mnras] {10.1093/mnras/stac974}, \href {https://ui.adsabs.harvard.edu/abs/2022MNRAS.513.4605P} {513, 4605}

\bibitem[\protect\citeauthoryear{{Perrone} \& {Latter}}{{Perrone} \& {Latter}}{2022b}]{Perrone2022b}
{Perrone} L.~M.,  {Latter} H.,  2022b, \mn@doi [\mnras] {10.1093/mnras/stac975}, \href {https://ui.adsabs.harvard.edu/abs/2022MNRAS.513.4625P} {513, 4625}

\bibitem[\protect\citeauthoryear{{Perrone}, {Berlok}  \& {Pfrommer}}{{Perrone} et~al.}{2024a}]{Perrone2024b}
{Perrone} L.~M.,  {Berlok} T.,   {Pfrommer} C.,  2024a, \mn@doi [arXiv e-prints] {10.48550/arXiv.2402.06718}, \href {https://ui.adsabs.harvard.edu/abs/2024arXiv240206718P} {p. arXiv:2402.06718}

\bibitem[\protect\citeauthoryear{{Perrone}, {Berlok}  \& {Pfrommer}}{{Perrone} et~al.}{2024b}]{Perrone2024a}
{Perrone} L.~M.,  {Berlok} T.,   {Pfrommer} C.,  2024b, \mn@doi [\aap] {10.1051/0004-6361/202347428}, \href {https://ui.adsabs.harvard.edu/abs/2024A&A...682A.125P} {682, A125}

\bibitem[\protect\citeauthoryear{{Pfrommer} \& {Dursi}}{{Pfrommer} \& {Dursi}}{2010}]{2010PfrommerDursi}
{Pfrommer} C.,  {Dursi} L.~J.,  2010, \mn@doi [Nature Physics] {10.1038/nphys1657}, \href {https://ui.adsabs.harvard.edu/abs/2010NatPh...6..520P} {6, 520}

\bibitem[\protect\citeauthoryear{{Pillepich} et~al.,}{{Pillepich} et~al.}{2018a}]{2018PillepichMethod+}
{Pillepich} A.,  et~al., 2018a, \mn@doi [\mnras] {10.1093/mnras/stx2656}, \href {https://ui.adsabs.harvard.edu/abs/2018MNRAS.473.4077P} {473, 4077}

\bibitem[\protect\citeauthoryear{{Pillepich} et~al.,}{{Pillepich} et~al.}{2018b}]{2018Pillepich+}
{Pillepich} A.,  et~al., 2018b, \mn@doi [\mnras] {10.1093/mnras/stx3112}, \href {https://ui.adsabs.harvard.edu/abs/2018MNRAS.475..648P} {475, 648}

\bibitem[\protect\citeauthoryear{{Planck Collaboration} et~al.,}{{Planck Collaboration} et~al.}{2020}]{2020Planck}
{Planck Collaboration} et~al., 2020, \mn@doi [\aap] {10.1051/0004-6361/201833910}, \href {https://ui.adsabs.harvard.edu/abs/2020A&A...641A...6P} {641, A6}

\bibitem[\protect\citeauthoryear{{Quataert}}{{Quataert}}{2008}]{2008Quataert}
{Quataert} E.,  2008, \mn@doi [\apj] {10.1086/525248}, \href {https://ui.adsabs.harvard.edu/abs/2008ApJ...673..758Q} {673, 758}

\bibitem[\protect\citeauthoryear{{Riquelme}, {Quataert}  \& {Verscharen}}{{Riquelme} et~al.}{2016}]{2016Riquelme+}
{Riquelme} M.~A.,  {Quataert} E.,   {Verscharen} D.,  2016, \mn@doi [\apj] {10.3847/0004-637X/824/2/123}, \href {https://ui.adsabs.harvard.edu/abs/2016ApJ...824..123R} {824, 123}

\bibitem[\protect\citeauthoryear{{Roberg-Clark}, {Drake}, {Reynolds}  \& {Swisdak}}{{Roberg-Clark} et~al.}{2016}]{2016Roberg-Clark+}
{Roberg-Clark} G.~T.,  {Drake} J.~F.,  {Reynolds} C.~S.,   {Swisdak} M.,  2016, \mn@doi [\apjl] {10.3847/2041-8205/830/1/L9}, \href {https://ui.adsabs.harvard.edu/abs/2016ApJ...830L...9R} {830, L9}

\bibitem[\protect\citeauthoryear{{Roberg-Clark}, {Drake}, {Reynolds}  \& {Swisdak}}{{Roberg-Clark} et~al.}{2018}]{2018Roberg-Clark+}
{Roberg-Clark} G.~T.,  {Drake} J.~F.,  {Reynolds} C.~S.,   {Swisdak} M.,  2018, \mn@doi [\prl] {10.1103/PhysRevLett.120.035101}, \href {https://ui.adsabs.harvard.edu/abs/2018PhRvL.120c5101R} {120, 035101}

\bibitem[\protect\citeauthoryear{{Ruszkowski} \& {Oh}}{{Ruszkowski} \& {Oh}}{2010}]{2010RuszkowskiOh}
{Ruszkowski} M.,  {Oh} S.~P.,  2010, \mn@doi [\apj] {10.1088/0004-637X/713/2/1332}, \href {https://ui.adsabs.harvard.edu/abs/2010ApJ...713.1332R} {713, 1332}

\bibitem[\protect\citeauthoryear{{Ruszkowski} \& {Oh}}{{Ruszkowski} \& {Oh}}{2011}]{2011RuszkowskiOh}
{Ruszkowski} M.,  {Oh} S.~P.,  2011, \mn@doi [\mnras] {10.1111/j.1365-2966.2011.18482.x}, \href {https://ui.adsabs.harvard.edu/abs/2011MNRAS.414.1493R} {414, 1493}

\bibitem[\protect\citeauthoryear{{Ruszkowski}, {Lee}, {Br{\"u}ggen}, {Parrish}  \& {Oh}}{{Ruszkowski} et~al.}{2011}]{2011Ruszkowski+}
{Ruszkowski} M.,  {Lee} D.,  {Br{\"u}ggen} M.,  {Parrish} I.,   {Oh} S.~P.,  2011, \mn@doi [\apj] {10.1088/0004-637X/740/2/81}, \href {https://ui.adsabs.harvard.edu/abs/2011ApJ...740...81R} {740, 81}

\bibitem[\protect\citeauthoryear{Saad \& Schultz}{Saad \& Schultz}{1986}]{1986SaadSchultz}
Saad Y.,  Schultz M.~H.,  1986, \mn@doi [SIAM Journal on Scientific and Statistical Computing] {10.1137/0907058}, 7, 856

\bibitem[\protect\citeauthoryear{{Schekochihin}, {Cowley}, {Kulsrud}, {Hammett}  \& {Sharma}}{{Schekochihin} et~al.}{2005}]{2005Schekochihin+}
{Schekochihin} A.~A.,  {Cowley} S.~C.,  {Kulsrud} R.~M.,  {Hammett} G.~W.,   {Sharma} P.,  2005, \mn@doi [\apj] {10.1086/431202}, \href {https://ui.adsabs.harvard.edu/abs/2005ApJ...629..139S} {629, 139}

\bibitem[\protect\citeauthoryear{{Schwarzschild}}{{Schwarzschild}}{1958}]{1958Schwarzschild}
{Schwarzschild} M.,  1958, {Structure and evolution of the stars.}

\bibitem[\protect\citeauthoryear{{Sedov}}{{Sedov}}{1959}]{1959Sedov}
{Sedov} L.~I.,  1959, {Similarity and Dimensional Methods in Mechanics}

\bibitem[\protect\citeauthoryear{{Sharma} \& {Hammett}}{{Sharma} \& {Hammett}}{2007}]{2007SharmaHammett}
{Sharma} P.,  {Hammett} G.~W.,  2007, \mn@doi [Journal of Computational Physics] {10.1016/j.jcp.2007.07.026}, \href {https://ui.adsabs.harvard.edu/abs/2007JCoPh.227..123S} {227, 123}

\bibitem[\protect\citeauthoryear{{Sharma} \& {Hammett}}{{Sharma} \& {Hammett}}{2011}]{2011SharmaHammett}
{Sharma} P.,  {Hammett} G.~W.,  2011, \mn@doi [Journal of Computational Physics] {10.1016/j.jcp.2011.03.009}, \href {https://ui.adsabs.harvard.edu/abs/2011JCoPh.230.4899S} {230, 4899}

\bibitem[\protect\citeauthoryear{{Sharma}, {Parrish}  \& {Quataert}}{{Sharma} et~al.}{2010a}]{2010Sharma+}
{Sharma} P.,  {Parrish} I.~J.,   {Quataert} E.,  2010a, \mn@doi [\apj] {10.1088/0004-637X/720/1/652}, \href {https://ui.adsabs.harvard.edu/abs/2010ApJ...720..652S} {720, 652}

\bibitem[\protect\citeauthoryear{{Sharma}, {Parrish}  \& {Quataert}}{{Sharma} et~al.}{2010b}]{2010Sharmab+}
{Sharma} P.,  {Parrish} I.~J.,   {Quataert} E.,  2010b, \mn@doi [\apj] {10.1088/0004-637X/720/1/652}, \href {https://ui.adsabs.harvard.edu/abs/2010ApJ...720..652S} {720, 652}

\bibitem[\protect\citeauthoryear{Smedt, Pattyn  \& Groen}{Smedt et~al.}{2010}]{2010Smedt+}
Smedt B.~D.,  Pattyn F.,   Groen P.~D.,  2010, \mn@doi [Journal of Glaciology] {10.3189/002214310791968395}, 56, 257–261

\bibitem[\protect\citeauthoryear{{Spitzer}}{{Spitzer}}{1962}]{1962Spitzer}
{Spitzer} L.,  1962, {Physics of Fully Ionized Gases}

\bibitem[\protect\citeauthoryear{{Springel}}{{Springel}}{2010}]{2010Springel}
{Springel} V.,  2010, \mn@doi [\mnras] {10.1111/j.1365-2966.2009.15715.x}, \href {https://ui.adsabs.harvard.edu/abs/2010MNRAS.401..791S} {401, 791}

\bibitem[\protect\citeauthoryear{{Springel} \& {Hernquist}}{{Springel} \& {Hernquist}}{2003}]{2003SpringelHernquist}
{Springel} V.,  {Hernquist} L.,  2003, \mn@doi [\mnras] {10.1046/j.1365-8711.2003.06206.x}, \href {https://ui.adsabs.harvard.edu/abs/2003MNRAS.339..289S} {339, 289}

\bibitem[\protect\citeauthoryear{{Springel} et~al.,}{{Springel} et~al.}{2018}]{2018Springel+}
{Springel} V.,  et~al., 2018, \mn@doi [\mnras] {10.1093/mnras/stx3304}, \href {https://ui.adsabs.harvard.edu/abs/2018MNRAS.475..676S} {475, 676}

\bibitem[\protect\citeauthoryear{{Springel}, {Pakmor}, {Zier}  \& {Reinecke}}{{Springel} et~al.}{2021}]{2021Springel+}
{Springel} V.,  {Pakmor} R.,  {Zier} O.,   {Reinecke} M.,  2021, \mn@doi [\mnras] {10.1093/mnras/stab1855}, \href {https://ui.adsabs.harvard.edu/abs/2021MNRAS.506.2871S} {506, 2871}

\bibitem[\protect\citeauthoryear{{Su} et~al.,}{{Su} et~al.}{2019}]{2019Su+}
{Su} K.-Y.,  et~al., 2019, \mn@doi [\mnras] {10.1093/mnras/stz1494}, \href {https://ui.adsabs.harvard.edu/abs/2019MNRAS.487.4393S} {487, 4393}

\bibitem[\protect\citeauthoryear{{Thomas}, {Pfrommer}  \& {Pakmor}}{{Thomas} et~al.}{2023}]{2023Thomas+}
{Thomas} T.,  {Pfrommer} C.,   {Pakmor} R.,  2023, \mn@doi [\mnras] {10.1093/mnras/stad472}, \href {https://ui.adsabs.harvard.edu/abs/2023MNRAS.521.3023T} {521, 3023}

\bibitem[\protect\citeauthoryear{{Tilley}, {Balsara}  \& {Howk}}{{Tilley} et~al.}{2006}]{2006Tilley+}
{Tilley} D.~A.,  {Balsara} D.~S.,   {Howk} J.~C.,  2006, \mn@doi [\mnras] {10.1111/j.1365-2966.2006.10747.x}, \href {https://ui.adsabs.harvard.edu/abs/2006MNRAS.371.1106T} {371, 1106}

\bibitem[\protect\citeauthoryear{{Vogelsberger}, {Sijacki}, {Kere{\v{s}}}, {Springel}  \& {Hernquist}}{{Vogelsberger} et~al.}{2012}]{2012Vogelsberger+}
{Vogelsberger} M.,  {Sijacki} D.,  {Kere{\v{s}}} D.,  {Springel} V.,   {Hernquist} L.,  2012, \mn@doi [\mnras] {10.1111/j.1365-2966.2012.21590.x}, \href {https://ui.adsabs.harvard.edu/abs/2012MNRAS.425.3024V} {425, 3024}

\bibitem[\protect\citeauthoryear{{Vogelsberger}, {Genel}, {Sijacki}, {Torrey}, {Springel}  \& {Hernquist}}{{Vogelsberger} et~al.}{2013}]{2013Vogelsberger+}
{Vogelsberger} M.,  {Genel} S.,  {Sijacki} D.,  {Torrey} P.,  {Springel} V.,   {Hernquist} L.,  2013, \mn@doi [\mnras] {10.1093/mnras/stt1789}, \href {https://ui.adsabs.harvard.edu/abs/2013MNRAS.436.3031V} {436, 3031}

\bibitem[\protect\citeauthoryear{{Voit}}{{Voit}}{2011}]{2011Voit}
{Voit} G.~M.,  2011, \mn@doi [\apj] {10.1088/0004-637X/740/1/28}, \href {https://ui.adsabs.harvard.edu/abs/2011ApJ...740...28V} {740, 28}

\bibitem[\protect\citeauthoryear{{Voit}, {Donahue}, {Bryan}  \& {McDonald}}{{Voit} et~al.}{2015}]{2015Voit+}
{Voit} G.~M.,  {Donahue} M.,  {Bryan} G.~L.,   {McDonald} M.,  2015, \mn@doi [\nat] {10.1038/nature14167}, \href {https://ui.adsabs.harvard.edu/abs/2015Natur.519..203V} {519, 203}

\bibitem[\protect\citeauthoryear{{Weinberger} et~al.,}{{Weinberger} et~al.}{2017}]{2017Weinberger+}
{Weinberger} R.,  et~al., 2017, \mn@doi [\mnras] {10.1093/mnras/stw2944}, \href {https://ui.adsabs.harvard.edu/abs/2017MNRAS.465.3291W} {465, 3291}

\bibitem[\protect\citeauthoryear{{Weinberger}, {Springel}  \& {Pakmor}}{{Weinberger} et~al.}{2020}]{2020Weinberger+}
{Weinberger} R.,  {Springel} V.,   {Pakmor} R.,  2020, \mn@doi [\apjs] {10.3847/1538-4365/ab908c}, \href {https://ui.adsabs.harvard.edu/abs/2020ApJS..248...32W} {248, 32}

\bibitem[\protect\citeauthoryear{{Yang} \& {Reynolds}}{{Yang} \& {Reynolds}}{2016}]{2016YangReynolds}
{Yang} H. Y.~K.,  {Reynolds} C.~S.,  2016, \mn@doi [\apj] {10.3847/0004-637X/818/2/181}, \href {https://ui.adsabs.harvard.edu/abs/2016ApJ...818..181Y} {818, 181}

\bibitem[\protect\citeauthoryear{{Ye}, {Shen}, {Lin}  \& {Mei}}{{Ye} et~al.}{2020}]{2020Ye+}
{Ye} J.,  {Shen} C.,  {Lin} J.,   {Mei} Z.,  2020, \mn@doi [Astronomy and Computing] {10.1016/j.ascom.2019.100341}, \href {https://ui.adsabs.harvard.edu/abs/2020A&C....3000341Y} {30, 100341}

\bibitem[\protect\citeauthoryear{{Yokoyama} \& {Shibata}}{{Yokoyama} \& {Shibata}}{1997}]{1997YokoyamaShibata}
{Yokoyama} T.,  {Shibata} K.,  1997, \mn@doi [\apjl] {10.1086/310429}, \href {https://ui.adsabs.harvard.edu/abs/1997ApJ...474L..61Y} {474, L61}

\bibitem[\protect\citeauthoryear{{Zakamska} \& {Narayan}}{{Zakamska} \& {Narayan}}{2003}]{2003ZakamskaNarayan}
{Zakamska} N.~L.,  {Narayan} R.,  2003, \mn@doi [\apj] {10.1086/344641}, \href {https://ui.adsabs.harvard.edu/abs/2003ApJ...582..162Z} {582, 162}

\bibitem[\protect\citeauthoryear{{Zel'dovich} \& {Raizer}}{{Zel'dovich} \& {Raizer}}{1967}]{1967ZeldovichRaizer}
{Zel'dovich} Y.~B.,  {Raizer} Y.~P.,  1967, {Physics of shock waves and high-temperature hydrodynamic phenomena}

\bibitem[\protect\citeauthoryear{{ZuHone}, {Kunz}, {Markevitch}, {Stone}  \& {Biffi}}{{ZuHone} et~al.}{2015}]{2015ZuHone}
{ZuHone} J.~A.,  {Kunz} M.~W.,  {Markevitch} M.,  {Stone} J.~M.,   {Biffi} V.,  2015, \mn@doi [\apj] {10.1088/0004-637X/798/2/90}, \href {https://ui.adsabs.harvard.edu/abs/2015ApJ...798...90Z} {798, 90}

\bibitem[\protect\citeauthoryear{van Leer}{van Leer}{1984}]{1984vanLeer}
van Leer B.,  1984, \mn@doi [SIAM Journal on Scientific and Statistical Computing] {10.1137/0905001}, 5, 1

\makeatother
\end{thebibliography}




\appendix
\section{Estimating the conduction coefficient in cell interfaces}
\label{app: avg}
\begin{figure}
    \centering
    \includegraphics[width=0.49\textwidth]{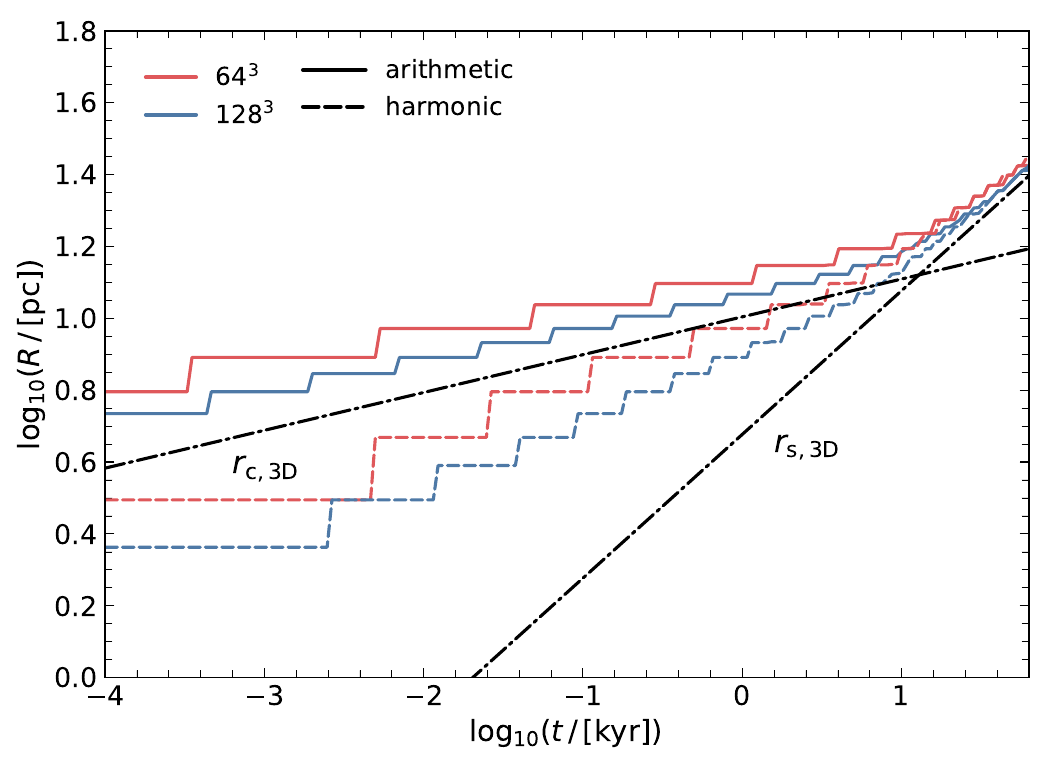}
    \caption[Sedov iso alt kappa]
    {Evolution of the radius of the conduction/shock front in the 3D point explosion test with isotropic conduction. Solid/dashed coloured lines show results from simulations where the conductivity in cell interfaces was calculated using the arithmetic/harmonic mean. The black, dot-dash lines show the analytic expectation for the radial evolution of the shock/conduction front in the cases of pure hydrodynamics ($r_{\rm s,3D}$, equation~\ref{eq: r iso ad}) and pure conduction ($r_{\rm c,3D}$, equation~\ref{eq: r iso cond}), respectively. With increasing resolution, the radius of the conduction front converges with the analytic solution when the conductivity is calculated using the arithmetic mean, and diverges from the analytics when the harmonic mean is used.}
    \label{fig: Sedov iso altkappa}
\end{figure}

In Section~\ref{subsub: interface} we explained that we calculate the conductivity in a cell interface by taking the arithmetic average of those of the two cells on each side of the interface (see equation~\ref{eq: arithmetic}). To check that this estimate is reasonable we also carried out isotropic point explosion tests (see Section~\ref{subsec: sedov}) where we estimated the conductivity in the interface by taking the harmonic average \citep[see e.g.][]{2007SharmaHammett} of those of the two cells on either side
\begin{equation}
\label{eq: harmonic}
    \chi_{ij,{\rm h}} = \frac{1}{\frac{1}{2}(\frac{1}{\chi_i} + \frac{1}{\chi_j})} \, .
\end{equation}
From equations~(\ref{eq: arithmetic}) and (\ref{eq: harmonic}), we can see that, in the case of significant differences in conductivity between the cells, the arithmetic mean will tend towards the larger value of $\chi$, while the harmonic mean tends towards the smaller.

We compare these two averaging procedures by carrying out simulations analogous to those presented in Section~\ref{subsub: sedov iso}, i.e.~in the isotropic blast wave setup. We do so because, in this setup, the speed of the conduction front can be predicted using analytical arguments. We show simulations at two different resolutions, $64^3$ and $128^3$, for each of these averaging procedures, which we will refer to as `arithmetic' and `harmonic'. All details of the simulations are as described in Section~\ref{subsec: sedov}. 

In Fig.~\ref{fig: Sedov iso altkappa} we show the time evolution of the radius of the shock/conduction in these simulations. Solid/dashed coloured lines show results from simulations where the conductivity in cell interfaces was calculated using the arithmetic/harmonic mean. From this figure it is clear that the speed of the conduction fronts in the `harmonic' case are slower than the analytic prediction, $r_{\rm c, 3D}$ and those measured in the `arithmetic' simulations. 

Additionally, as can be seen in Figures~\ref{fig: Sedov iso}~and~\ref{fig: Sedov iso altkappa} (and was discussed in Section~\ref{subsub: sedov iso}), with increasing resolution, the size of the conduction front is smaller (at a given time). In the `arithmetic' case, this means that the radius of the conduction front converges towards the analytic prediction, but for the `harmonic' case, the radius diverges further from the analytic solution. It is for this reason that we choose, in our solver, to calculate the interface conductivities via equation~(\ref{eq: arithmetic}).

From Fig.~\ref{fig: Sedov iso altkappa} we see that the late-time behaviour is similar in all simulations. This is due to the fact that hydrodynamical processes dominate at late times (see discussion in Section~\ref{subsub: sedov iso}). If this were not the case, we would expect that the differences in the sizes of the conduction fronts, seen at early times, would persist.


\bsp	
\label{lastpage}
\end{document}